\documentclass[aps,pra,superscriptaddress,preprint]{revtex4-1}
\usepackage{amsmath}
\usepackage{graphicx}
\usepackage{dcolumn}
\usepackage{bm}
\usepackage[export]{adjustbox}
\usepackage{subfig}
\usepackage{floatrow}
\newcommand{\mA}{m_\mathrm{A}}
\newcommand{\mB}{m_\mathrm{B}}
\newcommand{\mC}{m_\mathrm{C}}
\newcommand{\mD}{m_\mathrm{D}}

\newcommand{\mcX}{{\mathcal{X}}}

\begin{document}

\title{A quasiclassical method for calculating the density of states of ultracold collision complexes}
\author{Arthur Christianen}
\affiliation{Institute for Molecules and Materials, Radboud University, Nijmegen, The Netherlands}
\author{Tijs Karman}
\affiliation{\!ITAMP, \!Harvard-Smithsonian \!Center \!for \!Astrophysics, \! \!Cambridge, \!Massachusetts, \!02138, \!USA}

\date{\today}
\author{Gerrit C. Groenenboom}
\email{gerritg@theochem.ru.nl}
\affiliation{Institute for Molecules and Materials, Radboud University, Nijmegen, The Netherlands}

\begin{abstract}
	We derive a quasiclassical expression for the density of states (DOS)
of an arbitrary, ultracold, $N$-atom collision complex, for a general potential
energy surface (PES).  We establish the accuracy of our quasiclassical method
by comparing to exact quantum results for the K$_2$-Rb and NaK-NaK systems,
with isotropic model PESs. Next, we calculate the DOS for an accurate NaK-NaK
PES to be 0.124~$\mu$K$^{-1}$, with an associated Rice-Ramsperger-Kassel-Marcus
(RRKM) sticking time of 6.0~$\mu$s.  We extrapolate the DOS and sticking times
to all other polar bialkali-bialkali collision complexes by scaling with atomic
masses, equilibrium bond lengths, dissociation energies, and dispersion
coefficients.  The sticking times calculated here are two to three orders of
magnitude shorter than those reported by Mayle \emph{et al.}\ [Phys.\ Rev.\ A
{\bf 85}, 062712 (2012)]. We estimate dispersion coefficients and collision
rates between molecules and complexes.  We find that the sticking-amplified
three-body loss mechanism is not likely the cause of the losses observed in the
experiments.
\end{abstract}

\maketitle

\section{Introduction}

Ultracold dipolar gases have applications ranging from quantum computation
\cite{demille:2002, yelin:2006, ni:2018} and simulation of condensed matter
systems \cite{micheli:2006,buchler:2007,cooper:2009}, to controlled chemistry
\cite{krems:08,ospelkaus:2010b}, and high-precision measurements to challenge
the standard model \cite{andreev:2018}. Ultracold polar, bialkali gases in
their absolute ground state have been realized experimentally for nonreactive
species such as the bosonic $^{87}$Rb$^{133}$Cs \cite{takekoshi:2014,
molony:2014} and $^{23}$Na$^{87}$Rb \cite{guo:2016} molecules, and the
fermionic $^{23}$Na$^{40}$K \cite{park:2015,seesselberg:2018}.  The lifetime of
these molecules in the trap is less than a second for the bosonic species
\cite{takekoshi:2014,guo:2016} and a few seconds for $^{23}$Na$^{40}$K
\cite{park:2015}. The coherence time between hyperfine states of
$^{23}$Na$^{40}$K molecules has been shown to approach a second
\cite{park:2017}. There is potential for improving this further
\cite{park:2017} meaning that the trap lifetime of the molecules limits the
coherence time. Increasing the lifetime of these molecules is therefore pivotal
to realizing applications of these ultracold dipolar gases.

The mechanism limiting the lifetime is currently unknown, but it likely
involves ultracold collisions between the molecules \cite{park:2015,
park:2017,ye:2018}, which have been studied extensively in the literature
\cite{quemener:2012,mayle:2013}. In Refs.~\cite{ye:2018,gregory:2019} it is
shown that the loss is equally fast as in the case of reactive collisions and
that the diatom-diatom collisions are the rate determining step.  The current
hypothesis is that the loss mechanism involves the formation of long-lived
complexes of pairs of diatoms \cite{mayle:2013}.  These diatoms
have long sticking times because of their strong chemical interactions, which gives rise to a high density of states (DOS) and chaotic
dynamics.

Croft \emph{et al.}\ study ultracold reactive collisions for the triatomic
K$_2$+Rb system with converged quantum scattering calculations and report that
this required over $300\,000$ hours of CPU time \cite{croft:2017}. For
four-atom systems such as NaK+NaK, the computation time will even be orders of
magnitude larger, making such calculations unfeasible at this time. Mayle
\emph{et al.}~\cite{mayle:2012, mayle:2013} suggested using the
Rice-Ramsperger-Kassel-Marcus (RRKM) formalism \cite{levine:2005} to calculate
the sticking time $\tau$, \emph{i.e.}, the lifetime of those collision complexes, from the DOS $\rho$,
\begin{align}
	\label{eq:taustick}
        \tau = \frac{2\pi\hbar\rho}{N^{(0)}},
\end{align}
where $N^{(0)}$ is the number of states at the transition state.  The
transition state separates the collision complexes, treated classically, from
the pair of colliding molecules, treated quantum mechanically.  A surface
dividing the two regions can be chosen at some $R=R^{(0)}$, where $R$ is the
Jacobi scattering coordinate in the asymptotic region.  For ultracold
collisions of ground-state nonreactive molecules, there is asymptotically only
one open channel, $N^{(0)}=1$. To define the DOS, we may choose $R^{(0)}$ as
the smallest intermolecular distance at which $N^{(0)}=1$. However, in practice
the DOS already converges for smaller $R$ between 20 and 50~$a_0$.

The RRKM theory assumes ergodic dynamics.  This assumption was found to be
valid for the K+KRb sytem \cite{croft:2017a} and should also apply to strongly
interacting four-atom sytems that have an even higher DOS. Mayle \emph{et al.}\
use simple model potential energy surfaces (PESs), for which the DOS can be
calculated quantum mechanically.  However, their state counting contained an
error, explained in Sec.~\ref{sec:mayle}, that caused an overestimation of the
DOS.  Furthermore, their method is not applicable to realistic PESs that do
depend on the molecular orientation and vibrational coordinates.  Nevertheless,
their observation that the DOS of ultracold collision complexes is very large
and that the RRKM model is a useful tool to calculate the sticking times is
very valuable. Furthermore, the DOS is used as a parameter in multichannel quantum defect theory \cite{mayle:2012,mayle:2013}, which can be used to describe ultracold scattering. The only needed change in these calculations is to insert the corrected DOS.

Since the DOS is large, a quasiclassical calculation of the DOS is expected to
be accurate. In Sec.~\ref{sec:DOS} we derive a quasiclassical expression for
the DOS of an $N$-atom collision complex.  Our expression can be applied to
PESs that depend on the molecular orientation and vibrational coordinates.  In
Secs.\ \ref{sec:K2Rb} and \ref{sec:NaKNaK} we validate this method for
isotropic vibrational coordinate-independent PESs, such as used by Mayle
\emph{et al.}~\cite{mayle:2013}, which allows comparing to converged quantum
mechanical state counting.  In Sec.~\ref{sec:NaKNaK} we apply our method
to our recently calculated NaK-NaK PES \cite{christianen:2019} to accurately
compute the DOS of the NaK-NaK system.  In Sec.~\ref{sec:extrapol} we
extrapolate our results to also estimate the DOS for other polar bialkali
collision complexes.  Finally, we show in Sec.~\ref{sec:threebod} that the
sticking times are not large enough for a three-body loss mechanism to explain
the experimental losses.

\section{Theory}
\subsection{Counting angular momentum states \label{sec:mayle}}

To calculate the DOS quantum mechanically we count all quantum states in
a finite energy interval and divide by the size of the interval. Calculating
the quantum states is as difficult as solving the scattering problem, so
approximations are necessary.  In Ref.~\cite{mayle:2012} a method was developed
to count quantum states for three particle systems described by isotropic,
bond-length independent interaction potentials.  For such potentials, the DOS
calculation is simplified as the angular and vibrational coordinates are
uncoupled from one another, as well as from the intermolecular distance.
Hence, the DOS can be computed essentially by computing the DOS for the
one-dimensional radial problem and subsequently multiplying by the number of
contributing ro-vibrational states.

When counting states, it is important to take into account angular momentum
conservation, which is done most conveniently in a coupled representation.  For
atom-diatom systems, coupled states are denoted $|(jl) J M \rangle$,
where $j$ is the diatom rotational quantum number, $l$ corresponds
to the end-over-end angular momentum, $J$ to the total angular
momentum, and $M$ to the projection of the total angular momentum
on a space-fixed axis. Thus, for a given $J$ and $M$, there is
exactly one quantum state for each pair $(j,l)$ that satisfies
the triangular conditions $|j-J| \leq l \leq j+J$.
Mayle \emph{et al.}~\cite{mayle:2012}, however,
counted all uncoupled basis functions $|j m_j l m_l \rangle$ that have nonzero
overlap with specific $J$ (and $M=0$).  Hence, for each pair $j,l$ they count $2
\min(j,l)+1$ states, rather than one.  Because rotational states with $j$ in
the low hundreds can contribute energetically, this led to an overestimation of
the DOS by two to three orders of magnitude.  This has also been noted by Croft
\emph{et al.}~\cite{croft:2017} who corrected this mistake.

Furthermore, also the parity of the collision complex, $p=(-1)^{j+l}$, is
conserved. This means that we should only count the states with parity
$p=(-1)^{j_0+l_0}$, where $j_0$ and $l_0$ are the initial $j$ and $l$ of the
collision.  This constraint was not taken into account by either Mayle \emph{et
al.} \cite{mayle:2012,mayle:2013} or Croft \emph{et al.} \cite{croft:2017}.

In the presence of an external field, $J$ is no longer rigorously
conserved. In this case, the number of contributing states can be counted by
summing over $J$, which can lead to an increase of the DOS by
approximately four orders of magnitude.
When nonparallel electric and magnetic external fields are present,
also cylinder symmetry is broken, and $M$ is no longer conserved,
which can increase the DOS by approximately two more orders of magnitude.

The quantum mechanical state-counting method is limited to isotropic and bond
length-independent potentials. In the following section, we describe a
quasiclassical approach that is also applicable to more general potential
energy surfaces (PESs). The quasiclassical approach is accurate precisely
because the DOS is so large, meaning we are close to the classical limit.

\subsection{Quasiclassical DOS calculation} \label{sec:DOS}

In this section, we derive an expression for the DOS for an $N$-atom system with a general PES.
We compute the number of quantum states from the classical phase-space volume.
For a system of $N_i$ particles of type $i$, the total number of quantum states below a certain energy, $E$, with given total angular momentum $J_0$ and
center of mass (C.O.M.) $\bm{X}=(0,0,0)$ is given by
\begin{align}
  N^{(\mathrm{cl})}(E,{\bm J_0})=& \frac{1}{h^{3N-3} \prod_i{N_i!}} \int d{\bm x} \int d{\bm p}\ \theta\left[E-H(\bm{x},\bm{p})\right] \nonumber \\
&	\times \delta[{\bm P}({\bm p})]\ \delta[{\bm X}({\bm x})] \ \delta[{\bm J_0}-{\bm J}({\bm x},{\bm p})],
\end{align}
where $\theta(E)$ is the Heaviside step function, with $\theta(x)=0$ for $x<0$
and $\theta(x)=1$ for $x\ge0$. The factor $\ \prod_i{N_i!}$ corrects for indistinguishibility of the particles.  The DOS is the derivative of $N^{\mathrm{(cl)}}$
with respect to energy
\begin{align}
	\rho^{(\mathrm{cl})}(E,{\bm J_0}) &= \frac{d N^{(\mathrm{cl})} }{dE} = \frac{1}{h^{3N-3} \prod_i{N_i!}} \int d{\bm x} \int d{\bm p}\ \delta\left[E-H(\bm{x},\bm{p})\right] \nonumber \\
	\times& \delta[{\bm P}({\bm p})]\ \delta[{\bm X}({\bm x})]\ \delta[{\bm J_0}-{\bm J}({\bm x},{\bm p})].
\end{align}
The restrictions on the C.O.M. position, ${\bm X}({\bm x})$, and
momentum, ${\bm P}(\bm p)$, ensure their conservation as the C.O.M.
motion is uncoupled from the collision dynamics. Finally, the delta function in
${\bm J}({\bm x},{\bm p})$ restricts the classical total angular momentum.

The RRKM sticking time, Eq.~\eqref{eq:taustick}, scales with the DOS for
specific total angular momentum and projection quantum numbers, $J$ and $M$,
rather than the sharply defined classical angular momentum ${\bm J}$.
Therefore, we need to determine integration bounds for the classical total
angular momenta that correspond to the specific quantum numbers.
The relevant DOS can then be obtained by integrating over this subset of phase
space, which we denote symbolically as
\begin{align}
  \rho_{JMp}(E) = g_{\bm{N}Jp} \int_{JM} \rho^{(\mathrm{cl})}(E,{\bm J})\ d{\bm J},
\end{align}
where $p$ denotes the parity. Quantum mechanically the parity of the
total wavefunction is conserved during a molecular collision and is therefore a
good quantum number. Classically, this parity is not well defined, so a quantum
mechanical factor $g_{\bm{N},J,p}$ needs to be introduced, which is defined as the fraction of classical phase space with quantum numbers ${\bm{N}}$ and $J$ that is assigned parity $p$. This factor always obeys the relation: 
\begin{equation}
  1=g_{\bm{N},J,1}+g_{\bm{N},J,-1}.
\end{equation}
In some cases (see, e.g., Sec.~\ref{sec:K2Rb}), indistinguishability of the
atoms and angular momentum conservation restrict the parity, meaning that
$g_{\bm{N}Jp}=\delta_{p,1}$ or $g_{\bm{N}Jp}=\delta_{p,-1}$. In
most situations, however, half the DOS comes from even parity states
and the other half from odd parity states and 
$g_{\bm{N},J,p}$ approaches $\frac{1}{2}$ for both parities in
the limit of large rotational excitations of the collision partners.

For an arbitrary PES, we cannot analytically carry out the integrals over the
internal degrees of freedom on which the electronic energy depends.  However,
assuming the potential depends only on the coordinates, not the momenta, we can
carry out all integrals over momenta analytically.  This is complicated by the
restrictions on momentum and angular momentum conservation.  This means that we
need to switch to a coordinate system with the total angular momentum, and
C.O.M. position and momentum as coordinates.  We choose a coordinate
system with the minimal number of remaining integrals, which is the number of
internal coordinates $D=3N-6$.  Next, we need to determine the integration
bounds on the classical angular momenta.  Finally, we are in a position to
integrate over the momenta analytically.  The following subsections discuss
these three parts of the problem.

\subsubsection{Coordinate transformations}
We first transform the $3N$ Cartesian coordinates of the $N$ atoms,
$\{{\bm x}_i,\, i=1,\ldots, N\}$, to C.O.M. coordinates, $\bm X$, $zyz-$Euler angles for
the orientation of the complex, $\bm{\Omega}=(\alpha,\beta,\gamma)$, and
a set of $3N-6$ internal coordinates, $\bm{q}$, for which we use
Jacobi coordinates,
\begin{equation} \label{eq:coordtrafo}
\bm{x}_i = \bm{{X}} + \mathcal{R}(\bm{\Omega}) \bm{x}_i^{(\mathrm{bf})}(\bm{q}).
\end{equation}
The body-fixed coordinates $\bm{x}_i^{(\mathrm{bf})}$ are transformed to
space-fixed coordinates by the $3\times 3$ rotation matrix
$\mathcal{R}(\bm{\Omega})$.

The integrals over the delta functions in the C.O.M. position and momentum can now be carried out,
which leads to
\begin{equation}
  N_{JMp}(E) = g_{\bm{N}Jp} C_{Nm} \int_{JM}\!  d\bm{\Omega}\, d\bm{q}\,
  d\bm{\dot{\Omega}}\,  d\bm{\dot{q}}\; |\!\det{\mathcal{J}(\bm{q},\beta)}|^2\,
  \theta\left[E-H(\bm{\Omega},\bm{q},\bm{\dot{\Omega}},d\bm{\dot{q}})\right],
\end{equation}
where $\mathcal{J}(\bm{q},\beta)$ is the Jacobian matrix for the coordinate transformation of Eq.~\eqref{eq:coordtrafo}, which is independent of Euler angles $\alpha$ and $\gamma$, and
\begin{equation}
  C_{\bm{Nm}} =\frac{1}{h^{3N-3} (\sum_i N_i m_i)^3} \prod_i{\frac{m_i^{3N_i}}{N_i!}}. 
\end{equation} 

We replace the derivatives of the Euler angles by the angular momentum $\bm{L}$
associated with the rotation of the frame of the system. We use
$\bm{L}=\mathcal{I} \bm{\omega}$, where $\mathcal{I}$ is the inertial tensor of
the system and the angular velocity is
\begin{equation}
  \bm{\omega}=\begin{pmatrix}
  0 & -\sin \alpha & \cos \alpha \sin \beta \\
  0 & \cos \alpha & \sin \alpha \sin \beta \\
  1 & 0 &  \cos \beta 
  \end{pmatrix} \bm{\dot{\Omega}}.
\end{equation}
The Jacobian determinant for this transformation is given by $1/\sin{\beta}$.
This gives:
\begin{equation}
N_{JMp}(E) = g_{\bm{N}Jp}\, C_{\bm{Nm}} \int_{JM}\! d\bm{\Omega}\, d\bm{q}\,
d\bm{L}\, d\bm{\dot{q}}\; \frac{|\det{\mathcal{J}(\bm{q},\beta)}|^2}{\sin \beta
\det{\mathcal{I}(\bm{q})}}
\theta\left[E-H(\bm{\Omega},\bm{q},\bm{L},d\bm{\dot{q}})\right] .
\end{equation} 
We assume that the electronic energy depends only on the coordinates, $\bm{q}$,
and not their derivatives. This means the above integral can be separated as
\begin{equation} \label{eq:NsthE}
  N_{JMp}(E) = g_{\bm{N}Jp}\, C_{\bm{Nm}} \int \! d\bm{\Omega}\, d\bm{q}\,
  \frac{|\det{\mathcal{J}(\bm{q},\beta)}|^2}{\sin \beta
  \det{\mathcal{I}(\bm{q})}} 
  \int_{JM}\! d\bm{L}\,  d\bm{\dot{q}}
  \; \theta\left[E-V(\bm{q})-T_\mathrm{kin}(\bm{\Omega},\bm{q},\bm{L},\bm{\dot{q}})\right].
\end{equation}

During a molecular collision, the total angular momentum $\bm{J}$ is conserved.
To impose this restriction, we replace the integral over $\bm{L}$ by an
integral over $\bm{J} = \bm{L}+\mathcal{R}(\bm{\Omega}) \bm{j}$.  Here,
$\bm{j}=\sum_i m_i \bm{x}_i^{(\mathrm{bf})}(\bm{q}) \times
\bm{\dot{x}_i}^{(\mathrm{bf})}(\bm{q},\bm{\dot{q}}) $ is the angular momentum
in the body-fixed frame, often called the ``vibrational angular momentum'', and
$\bm{\dot{x}_i}^{(\mathrm{bf})}(\bm{q},\bm{\dot{q}})=\mathcal{K}_i(\bm{q})
\bm{\dot{q}}$, with $[\mathcal{K}_i]_{jk}=\partial [\bm{x}_i]_j/\partial
\bm{q}_k$.  The Jacobian determinant for the transformation from $\bm{L}$ to $\bm{J}$ is
unity. 

Furthermore, we need an expression for the kinetic energy
$T_\mathrm{kin}(\bm{\Omega},\bm{q},\bm{J},\bm{\dot{q}})$.  The time derivative of
coordinates in the space-fixed coordinate system $\bm{\dot{x}}_i$ can be
written in terms of the coordinates in the body-fixed frame as
\begin{equation}
  \bm{\dot{x}}_i=\bm{\dot{X}}+ \bm{\omega} \times \mathcal{R}(\bm{\Omega})
  \bm{x}_i^{(\mathrm{bf})}(\bm{q}) + \mathcal{R}(\bm{\Omega})
  \bm{\dot{x}}_i^{(\mathrm{bf})}(\bm{q},\bm{\dot{q}}).
\end{equation}
In the C.O.M. frame, the total kinetic energy can now be written as
\begin{multline}
  T_\mathrm{kin}=\sum_i \frac{m_i}{2} \biggl\{\left[\bm{\omega} \times
  \mathcal{R}(\bm{\Omega}) \bm{x}_i^{(\mathrm{bf})}(\bm{q})\right]^2 +
  \left[\bm{\dot{x}_i}^{(\mathrm{bf})}(\bm{q},\bm{\dot{q}})\right]^2  + \\ 2
  \left[\bm{\omega} \times
  \mathcal{R}(\bm{\Omega})\bm{x}_i^{(\mathrm{bf})}(\bm{q},\bm{\dot{q}})\right]
  \cdot \mathcal{R}(\bm{\Omega}) \bm{\dot{x}}^{(\mathrm{bf})}_i
  (\bm{q},\bm{\dot{q}}) \biggl\}.
\end{multline}
We define the inertial tensor, $\mathcal{I}^{(\mathrm{bf})}(\bm{q})$, in the
body-fixed frame, and we use
\begin{equation}
  \bm{L}=\mathcal{R}(\bm{\Omega})\mathcal{I}^{(\mathrm{bf})}(\bm{q})
  \mathcal{R}(\bm{\Omega})^{-1}
  \bm{\omega}=\bm{J}-\mathcal{R}(\bm{\Omega})\bm{j}. 
\end{equation}
This yields
\begin{equation}
  T_\mathrm{kin}=\sum_i \frac{m_i
  [\bm{\dot{x}}_i^{(\mathrm{bf})}(\bm{q},\bm{\dot{q}})]^2}{2}-\frac{\bm{j}^T
  [\mathcal{I}^{(\mathrm{bf})}(\bm{q})]^{-1} \bm{j}}{2}+\frac{\bm{J}^T
  \mathcal{R}(\bm{\Omega}) [\mathcal{I}^{(\mathrm{bf})}(\bm{q})]^{-1}
  \mathcal{R}(\bm{\Omega})^{-1} \bm{J}}{2}. 
\end{equation}
We can write
$\bm{\dot{x}}_i^{(\mathrm{bf})}(\bm{q},\bm{\dot{q}})=\mathcal{K}_i(\bm{q})
\bm{\dot{q}}$ and $\bm{j}=\sum_i \mathcal{D}_i \bm{\dot{q}}$, with
\begin{equation}
 [\mathcal{D}_i]_{jk}=m_i \sum_{lm} \epsilon_{jlm} [\bm{x}_i^{(\mathrm{bf})}
(\bm{q})]_l [\mathcal{K}_i(\bm{q})]_{mk}, 
\end{equation}

where $\epsilon_{ijk}$ is the Levi-Civita tensor.  Therefore, we can generally
write the kinetic energy as a quadratic form in $\bm{\dot{q}}$,
\begin{equation}
  T_\mathrm{kin}(\bm{\Omega},\bm{q},\bm{J},\bm{\dot{q}})=\bm{\dot{q}}^T
  \mathcal{A}(\bm{q}) \bm{\dot{q}}+\frac{\bm{J}^T \mathcal{R}(\bm{\Omega})
  [\mathcal{I}^{(\mathrm{bf})}(\bm{q})]^{-1} \mathcal{R}(\bm{\Omega})^{-1}
  \bm{J}}{2},
\end{equation}
where $\mathcal{A}(\bm{q})$ is given by
\begin{equation}
  \mathcal{A}(\bm{q})=\frac{1}{2}\sum_i \mathcal{K}_i(\bm{q})^T m_i \mathcal{K}_i(\bm{q}) -
  \mathcal{D}_i(\bm{q})^T [\mathcal{I}^{(\mathrm{bf})}(\bm{q})]^{-1}
  \mathcal{D}_i(\bm{q}).
\end{equation}
Substituting this into Eq.~\eqref{eq:NsthE} gives
\begin{align}  \label{eq:nstates}
  N_{JMp}(E)=& g_{\bm{N}Jp} C_{\bm{Nm}}\int_{JM}\! d\bm{\Omega}\, d\bm{q}\,
   d\bm{J}\, d\bm{\dot{q}}\;
   \frac{|\det{\mathcal{J}(\bm{q},\beta)}|^2}{ \sin \beta
  \det{\mathcal{I}(\bm{q})}} \nonumber \\
  &\times \theta\left[E-V(\bm{q})-\bm{\dot{q}}^T \mathcal{A}(\bm{q}) \bm{\dot{q}}
  -\frac{\bm{J}^T \mathcal{R}(\bm{\Omega})
  [\mathcal{I}^{(\mathrm{bf})}(\bm{q})]^{-1} \mathcal{R}(\bm{\Omega})^{-1}
  \bm{J}}{2} \right].
\end{align}

\subsubsection{Angular momentum integration bounds}
The next step is to  determine the
integration range for the classical vector $\bm{J}$ that corresponds to a
specific quantum number $J$. Quantum mechanically we count the states
$|(jL)JM\rangle$.  The values $j$ can reach are typically very large for the
strongly interacting systems we consider here \cite{mayle:2012}, and we are
interested in ultracold collisions, meaning $J$ is small. Therefore we use the
approximation $j > J$ so that for each value of $j$, there are $2J+1$
allowed values of $L$ for each pair of quantum numbers $J$ and $M$.  Since
space is isotropic, the DOS does not depend on $M$.  Therefore, we
integrate over all phase-space regions corresponding to the allowed
$M$-values for the given $J$ and subsequently divide by $2J+1$.  The total
integral over the region corresponding to quantum number $J$ scales as
$(2J+1)^2$. Classically this is associated with the three dimensional integral
over the total angular momentum vector, $\bm{J}$,
\begin{equation}
  \int_J d\bm{J}=4 \pi \int_{B_J}^{B_{J+1}} |\bm{J}|^2 d|\bm{J}| = \frac{4}{3}
  \pi (B_{J+1}^3-B_{J}^3),
\end{equation}
where $B_J$ is the lower integration boundary of the classical region that
corresponds to the quantum number $J$. It is not directly evident what those
boundaries should be. However, we know the integral should be proportional to
$(2J+1)^2$. We can therefore derive a recurrence relation
\begin{equation}
  B_{J+1}^3-B_{J}^3=\left(\frac{2J+1}{2J-1}\right)^2 (B_{J}^3-B_{J-1}^3).
\end{equation}
This recursion relation can be solved with $B_0=0$ to yield
\begin{equation}
  B_J=[\frac{1}{3} J(2J-1)(2J+1)]^{\frac{1}{3}} B_1.
\end{equation}
If $J \gg 1$ then this expression approaches
\begin{equation}
  B_{J}=(\frac{4}{3})^{\frac{1}{3}} J B_1.
\end{equation}
Because the angular momentum at quantum number $J$ is given by $\sqrt{J (J+1)}
\hbar \rightarrow (J+\frac{1}{2}) \hbar$, the expression for $B_J$ should go to
$\hbar J$.  This means that $B_1 = \frac{3}{4}^{\frac{1}{3}} \hbar$. This value
of $B_1$ is also consistent with the quasiclassical quantization, since the
integral over $\bm{J}$ and its conjugate variable, $\bm{\Omega}$ should give
$h^3$ for $J=0$.
The integral over $\bm{J}$ with the given value of $B_1$ yields $\pi \hbar^3$.
If this is combined with the integral over $\bm{\Omega}$, which gives $8
\pi^2$, we obtain $8 \pi^3 \hbar^3=h^3$, as expected.

\subsubsection{Carrying out the integration}
Given the integration range corresponding to the total angular momentum
$\bm{J}$ we can carry out the integral of Eq.~\eqref{eq:nstates}. In the
ultracold regime, without an external field breaking angular momentum
conservation, $J$ is very small and the energy term $\bm{J}^T
\mathcal{R}(\bm{\Omega}) [\mathcal{I}^{-1}]^{(\mathrm{bf})} \mathcal{R}(\bm{\Omega})^{-1} \bm{J}$
is negligible compared to the interaction energy $V(\bm{q})$. The integral over
$\bm{J}$ will therefore yield a constant value of $\pi (2J+1) \hbar^3$ and the
integrand no longer depends on $\bm{\Omega}$. If the integration over $\bm{J}$
and $\bm{\Omega}$ is carried out, the following expression remains
\begin{equation} 
  N_{JMp}(E)=g_{\bm{N}Jp}\, 8 \pi^3 (2J+1)\hbar^3 C_{\bm{Nm}} \int \! d\bm{q}\,
  \frac{|\det{\mathcal{J'}(\bm{q})}|^2}{\det{\mathcal{I}(\bm{q})}} 
  \int\! d\bm{\dot{q}}\, \theta\left[E-V(\bm{q})-\bm{\dot{q}}^T \mathcal{A}(\bm{q})
  \bm{\dot{q}}\right],
\end{equation} 
where $\mathcal{J'}(\bm{q})=\mathcal{J}(\bm{q},\bm{\beta})/\sin \beta$.
Note that $\mathcal{J}$ contains one factor $\sin \beta$. The matrix
$\mathcal{A}$ is positive definite such that the integral over $\bm{\dot{q}}$
is the volume of a hyperellipsoid. Therefore, with $D$ the dimension of
$\bm{q}$, the resulting expression is
\begin{equation} \label{eq:genPSV}
  N_{JMp}(E)=\frac{g_{\bm{N}Jp}\, 8 \pi^{3+\frac{D}{2}} (2J+1) \hbar^3
  C_{\bm{Nm}}}{\Gamma(\frac{D}{2}+1)} 
  \int\! d\bm{q}\, G(\bm{q})\, [E-V(\bm{q})]^{\frac{D}{2}}{}.
\end{equation} 
We call the factor
\begin{equation}
G(\bm{q})=\frac{|\det{\mathcal{J'}(\bm{q})}|^2}{\det{\mathcal{I}(\bm{q})}\sqrt{\det{\mathcal{A}(\bm{q})}}}
\end{equation}
the geometry factor.  The DOS of the system, $\rho = dN/dE$ is given by
\begin{equation} \label{eq:genDOS}
  \rho_{JMp}(E)=\frac{g_{\bm{N}Jp}\, 8 \pi^{3+\frac{D}{2}} \hbar^3 C_{\bm{Nm}}
  (2J+1)}{\Gamma(\frac{D}{2})}  \int\! d\bm{q}\, G(\bm{q})\,
  [E-V(\bm{q})]^{\frac{D}{2}-1} .
\end{equation} 
In general, this integral has to be evaluated numerically.

\subsubsection{The DOS in presence of external fields}

Above we considered the case where $J$, $M$, and $p$ are rigorously conserved,
as is the case for any collisional complex in the absence of external fields.
However, in the presence of a single external field, the Hamiltonian has
cylindrical symmetry, such that $J$ and $p$ are no longer conserved, but $M$
still is.  If multiple external fields---say electric and magnetic---occur at an
angle to one another, the cylindrical symmetry is also broken, and neither $J$
nor $M$ is rigorously conserved.  In the limit of strong fields, all values of
$J$ (and $M$) can be populated, whereas in the limit of weak fields, $J$ and
$M$ are conserved as discussed above.  For intermediate field strengths, the
coupling between the different $J$ states is small, meaning that the full
parameter space may not be explored within the sticking time. The statistical
theory assumes ergodicity but does not quantify the field strength at which the dynamics becomes ergodic in $J$. Purely statistically, we can only treat the strong (or zero) field
limit.  We assume that even in the strong field limit the interaction of the
molecules with the field is small compared to the interaction between the
molecules.

When both $J$ and $M$ are not conserved, the phase-space integral is easier
than in the case without a field, because we can treat the integration over
$\bm{J}$ the same as the integration over $\bm{\dot{q}}$.  This leads to a
factor $\sqrt{\det[\mathcal{I}(\bm{q})^{-1}/2]}$ in the denominator of
Eq.~\eqref{eq:genPSV} (note that the label (bf) is dropped, since the determinant is invariant under rotation) and an increase of the exponent of the energy by
$3/2$, yielding
\begin{equation} \label{eq:DOS_noJM}
  \rho(E)=\frac{16 \sqrt{2} \pi^3 C_{\bm{Nm}} }{\Gamma(\frac{D}{2}+\frac{3}{2})}
    \int\!  d\bm{q}\, G(\bm{q})\sqrt{\det{\mathcal{I}(\bm{q})}}\; \{\pi [E-V(\bm{q})
  ]\}^{\frac{D}{2}+\frac{1}{2}}. 
\end{equation}

In case only $J$ is not conserved, but $M$ still is, the integral is more
difficult since then $M$ introduces directionality in space. The derivation for
this case is given in the Appendix Sec.~\ref{appendix}.  The result is
\begin{equation} \label{eq:DOS_noJ}
  \rho(E)=\frac{16 \pi^3 C_{\bm{Nm}}}{\Gamma(\frac{D}{2}+1)} \int\!  d\bm{q}\,
  G(\bm{q})\,
  \sqrt{\frac{\det{\mathcal{I}(\bm{q})}}{\mathcal{I}_\mathrm{rot}(\bm{q})}}\;
  \{\pi [E-V(\bm{q}) ]\}^{\frac{D}{2}},
\end{equation}
where $\mathcal{I}_\mathrm{rot}$ is defined by a series expansion in Appendix
Sec.~\ref{appendix} and can be interpreted as a weighted average of the
eigenvalues of $\mathcal{I}$.

\section{Results}

First, we establish the validity of our quasiclassical approach by considering
simple model potentials for K$_2$-Rb and NaK-NaK for which quantum calculations
of the DOS are possible. Then we calculate the DOS for a realistic PES and use
this result to estimate the DOS for other alkali dimer complexes. We assume
for both K$_2$Rb and NaK-NaK that all identical atoms are in the same hyperfine
state, meaning they are indistinguishable. If the sticking time is long enough
for transitions between hyperfine states to occur during collisions, the DOS
increases not just by a factor corresponding to the number of hyperfine states,
but because the hyperfine angular momentum couples with the rotational angular
momentum, also higher $J$ and $M$ states become accessible, leading to an
increase of the DOS by orders of magnitude.

\subsection{K$_2$-Rb} \label{sec:K2Rb}

For a three-atom system, the geometry factor $G(\bm{q})$ is a simple expression
in terms of Jacobi coordinates, $\bm{q}=(R,r,\theta)$. Here, $R$ is the
distance beween Rb and the C.O.M. of K$_2$, $r$ is the bond length of the diatom and $\theta$ is
the polar angle. For a general three-atom system (AB+C), the expression for the
field-free DOS becomes
\begin{equation} \label{eq:DOS_AB+C}
  \rho^{(\mathrm{AB}+\mathrm{C})}_{\bm{N}Jp}(E)=\frac{{g_{\bm{N}Jp}}\,4 \sqrt{2}\, \pi(2J+1) \mA
  \mB \mC}{h^3 (\mA+\mB+\mC) g_\mathrm{ABC}} \int\! \frac{R r}{\sqrt{\mu
  R^2+\mu_\mathrm{AB} r^2}}  [E-V(\bm{q})]^{\frac{1}{2}} dR\, dr\, d\theta.
\end{equation}
Here, $g_\mathrm{ABC}=\prod_i{N_i!}$ is a degeneracy factor to account for
indistinguishability, $\mu=(\mA+\mB) \mC/(\mA+\mB+\mC)$ is the
reduced mass of the three-body system and
$\mu_\mathrm{AB}=\mA \mB/(\mA+\mB)$ is
the reduced mass of the diatom.  This agrees with expressions in the literature
for three-body systems, for example Al$_3$ \cite{peslherbe:1994}, except for
the degeneracy factor $g_\mathrm{ABC}$ and the parity dependent
factor $g_{{\bm N}Jp}$ which were not taken into account there. 

We use K$_2$-Rb as a model system, for which A and B are K and C is Rb.
Expressions for the kinetic energy, inertial tensor and the body-fixed angular
momentum $\bm{j}$ are given in appendix \ref{appendix2}.  To test the quality
of the quasiclassical approximation we use an isotropic, $r$-independent
Lennard-Jones interaction potential, as in Refs.~\cite{mayle:2012, mayle:2013,
croft:2017}, such that the potential energy is given by
\begin{equation}
  V(R,r,\theta)=\frac{C_{12}}{R^{12}}-\frac{C_6}{R^6}+V_{\mathrm{K}_2}(r).
\end{equation}
Here, $C_{12}=C_6^2/(4D_e)$ and $C_6$ are the Lennard-Jones parameters, and
$V_{\mathrm{K}_2}$ is the diatom potential of K$_2$. We use $C_6=8599~E_h
a_0^6$ and $D_e=1630$ cm$^{-1}$, which are twice the values of the K-Rb
potential. This is the same potential as used by Croft \emph{et
al.}~\cite{croft:2017}, except that we use for $V_{\mathrm{K}_2}$ the diatom
potential constructed for our previous work in Ref. \cite{christianen:2019}.
Just as in Ref.\ \cite{croft:2017} we only take into account even $j$ to
account for the indistinguishability of the K-atoms. If $j$ is even and
$J=0$, then $l=n$ and $p=1$, therefore $g_{\bm{N}Jp}=\delta_{p,1}$. 

For such an isotropic $r$-independent PES it is possible to converge the DOS
quantum mechanically and to compare this to our quasiclassical results.  In
the quasiclassical calculation, the remaining integrals in
Eqs.~\eqref{eq:genDOS}, \eqref{eq:DOS_noJM}, and \eqref{eq:DOS_noJ}, were
computed numerically.  The numerical integration was done using an integration
grid of 56 equidistant points in $r$ ranging from 3.5 to 9 $a_0$, 171 equidistant
points in $R$ ranging from 3 to 20 $a_0$, and 4 points in $\theta$ placed on a
Gauss-Legendre quadrature. The large grids in $r$ and $R$ are needed to
converge the low energy results.  To find the DOS quantum mechanically, we exploit the separation of radial and ro-vibrational degrees of freedom permitted by the isotropic $r$-independent PES. We compute the DOS for the one-dimensional radial problem, subsequently multiply by the number of contributing ro-vibrational states, and finally determine the DOS by binning the quantum states in an interval of 10 cm$^{-1}$ and divide their number by the interval length.

\begin{figure}
\begin{center}
\includegraphics[height=2.4in]{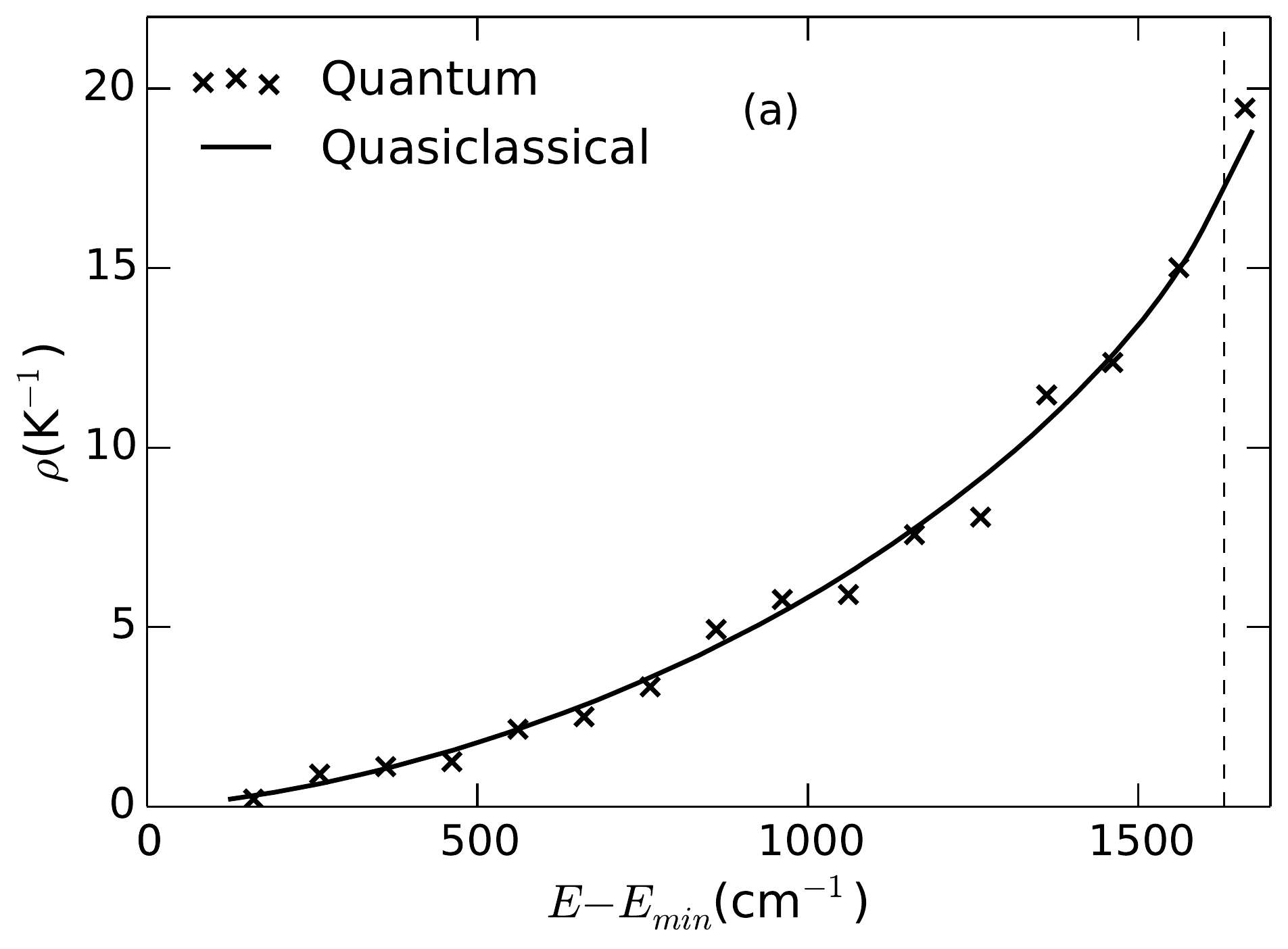}\\
\includegraphics[height=2.4in]{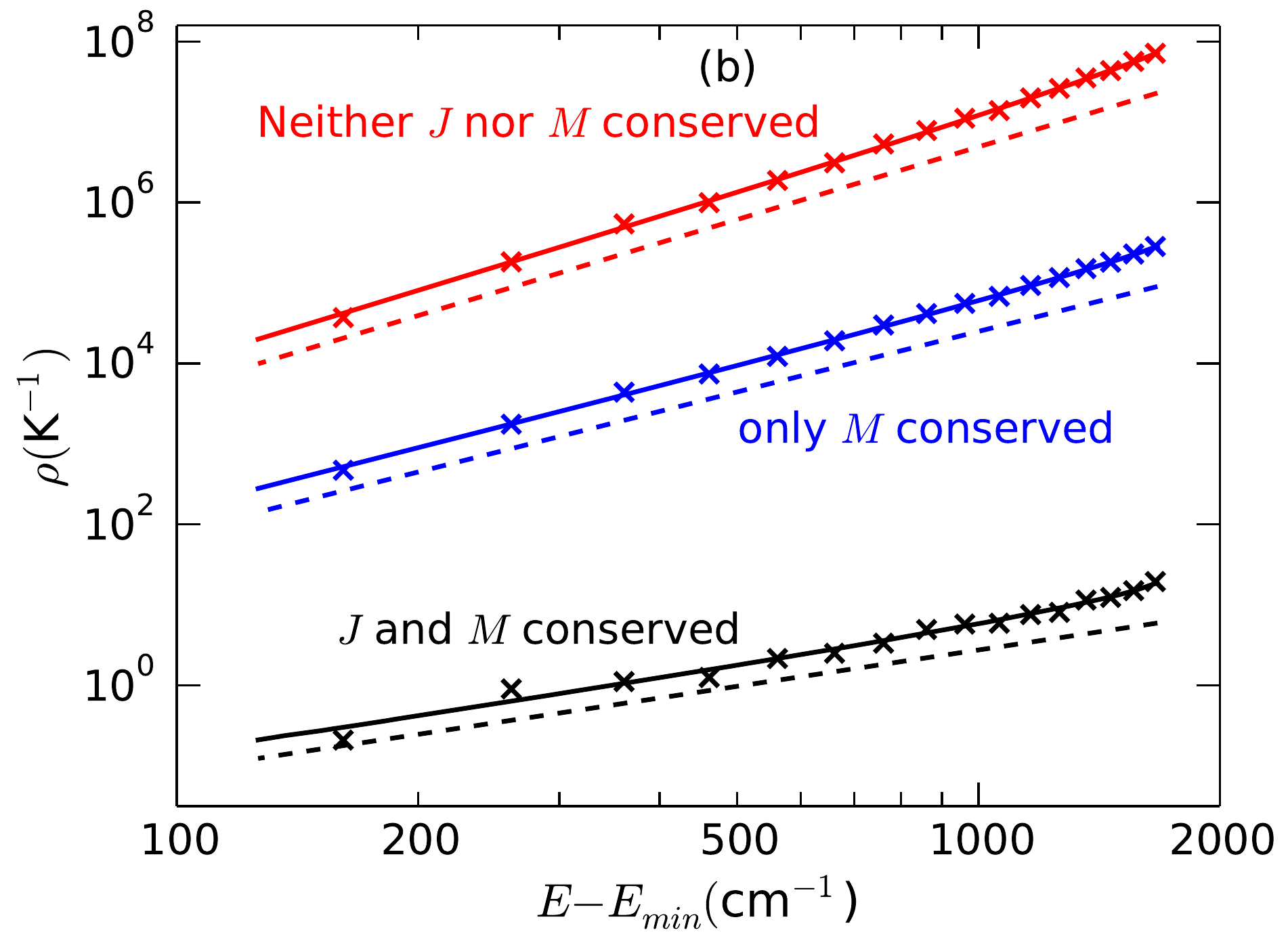}
\captionsetup{format=plain,justification=raggedright,singlelinecheck=false}
\caption{ \label{fig:fig1}
  The DOS of K$_2$+Rb as a function of the energy, $E$, for quantum mechanical
  (crosses) and quasiclassical (solid line) calculations for an isotropic PES. In panel (a) this is
  shown for the field-free case on a linear scale, in panel (b) it is also
  plotted for the cases with fields on a double logarithmic scale.
  The vertical dashed line in panel (a) indicates the classical dissociation limit
  of the complex with $E-E_\mathrm{min}=D_e$.
  The dashed lines in panel (b) are straight lines with slopes 3, 2.5, and 1.5
  and only serve to illustrate the
  power-law energy dependence of the DOS. Note that the crosses in the upper and
  lower graph are placed at the same energies.
}
\end{center}
\end{figure}

In Fig.~\ref{fig:fig1}(a) we show the DOS as a function of the energy
$E-E_\mathrm{min}$, where $E_\mathrm{min}$ is the energy of the minimum of the
potential. The vertical dashed line indicates the classical dissociation
limit for formation of Rb+K$_2$.
Quantummechanically, the dissociation energy liest at slightly higher energy
because of the zero-point energy of K$_2$. To compute the classical
DOS we place the dividing surface at $R=R^{(0)}=20~a_0$.
Above the classical dissociation limit the DOS keeps increasing 
when we move the dividing surface outwards, but only slowly.

The classical and quantum results agree closely with each other, especially
below the classical dissociation limit. In the quantum case there are more
fluctuations in the DOS, as expected. In Fig.~\ref{fig:fig1}(b) the DOSs are
plotted on a double logarithmic scale, both for the case without field and with
field(s).  Again, the classical and quantum mechanical results agree very well.
In the quantum case, the fluctuations become smaller as the DOS becomes larger.

The DOSs in \ref{fig:fig1}(b) show a clear power-law dependence on the energy,
$E$.  Straight dashed lines with slopes from top to bottom 3, 2.5, and 1.5 are
plotted alongside the DOSs to guide the eye.  The integrand of
Eq.~\eqref{eq:DOS_AB+C} has exponent $1/2$, and when $J$ or $J$ and $M$
are not conserved these exponents become $3/2$ and 2, respectively.
Here, we use an isotropic potential, which therefore does not depend on
$\theta$. Each ``harmonic'' degree of freedom contributes $1/2$ to the
exponent, such that the exponent would increase by one if the potentials as a
function of $R$ and $r$ were perfectly harmonic.  The slopes of the graphs are
slightly higher near dissociation.

\subsection{NaK-NaK} \label{sec:NaKNaK}

Next, we apply our method to the four-atom NaK-NaK system. We use the Jacobi
coordinates $\bm{q}=(R, r_1, r_2, \theta_1, \theta_2, \phi)$, where $R$ is the
NaK-NaK distance, $r_{1}$ and $r_2$ are the bond lengths, $\theta_{1}$ and
$\theta_2$
the polar angles, and $\phi$ is the dihedral angle. In these coordinates,
Eq.~\eqref{eq:genDOS} for the general AB+CD DOS can be written as
\begin{equation}\label{eq:DOS_ABCD}
  \rho_{JMp}^{(\mathrm{AB}+\mathrm{CD})}(E)=\frac{g_{\bm{N}Jp}\, 4 \pi^{6} (2J+1) \mA^3 \mB^3
  m_C^3 m_D^3}{h^{9} (\mA+\mB+\mC+\mD)^3 g_\mathrm{ABCD}}
   \int\! \frac{R^4\, r_1^4\,r_2^4
  \sin^2(\theta_1)\sin^2(\theta_2)}{\det{\mathcal{I}(\bm{q})}
  \sqrt{\det{\mathcal{A}(\bm{q})}}}  [E-V(\bm{q})]^{2}\, d\bm{q}.
\end{equation}
Unlike for the three-atom system, there is no simple analytical expression for
$\det{\mathcal{I}(\bm{q})}$ and $\det{\mathcal{A}(\bm{q})}$ for the four-atom
system, so we calculated them numerically.  For the NaK-NaK system, A and C are
K, B and D are Na, and $g_{\bm{N}Jp}=1/2$.  The expressions for
$\mathcal{I}$ and $\bm{j}$ are given in appendix \ref{appendix2}. In the
quasiclassical calculations for the isotropic PES, we use an equidistant grid
in $R$ from 5 to 20 $a_0$ with 151 points, a grid of $r_1$ from 4.5 to
10 $a_0$ with 56 points, a grid of $r_2$ ranging from $r_1$ to $10.5~a_0$ with
a spacing of 0.1 $a_0$. We use a four-point Gauss-Legendre quadrature in
$\theta_1$ and $\theta_2$ and a two-point Gauss-Chebyshev quadrature in $\phi$.
We choose $r_2>r_1$ and multiply the result by a factor two because of the
symmetry. An additional factor two is included to compensate for $\phi$ from
running up to $\pi$ instead of $2 \pi$.

The realistic potential energy surface of NaK-NaK consists of three
parts \cite{christianen:2019}: two symmetrically equivalent NaK-NaK parts and
one Na$_2$-K$_2$ part. Although one set of Jacobi coordinates can in principle
describe all arrangements, integrating over these Jacobi
coordinates is very difficult, because an increasingly fine angular
grid is needed when going further into an arrangement that does not match the
chosen coordinates. We therefore construct a separate integration grid in
Jacobi coordinates for all three arrangements and add the integrals. In the
NaK-NaK arrangement for the realistic potential, we use an equidistant grid in
$R$ with 31 points placed from 5 to 20 $a_0$. For $r_1$ we use a grid of 15
points from 4.5 to $9~a_0$, and for $r_2$ the grid ranges from $r_1$ to
$10.5~a_0$, with a spacing of $0.3~a_0$. A 24-point Gauss-Legendre quadrature
between 0 and $\pi$ is used for $\theta_1$ and $\theta_2$, and an 8-point
Gauss-Chebyshev quadrature between 0 and $\pi$ is used for $\phi$. For the
Na$_2$-K$_2$ we use a similar grid.

Because there are some overlapping parts of the grids in the center of the PES,
we assign a geometry dependent weighting factor to the integrands for each
arrangement. This weighting factor $W(\bm{q})$ is based on the symmetrization
function in our previous work \cite{christianen:2019}. In the NaK-NaK
arrangements $W(\bm{q})=W_1 W_2$ or $W(\bm{q})=W_1 (1-W_2)$, and in the
Na$_2$-K$_2$ arrangement: $W(\bm{q})=1-W_1$, with
\begin{equation}
  W(u,c,w)=\begin{cases}
  0, & \quad \text{if } u \leq c-w \\
  \frac{1}{2}+\frac{9}{16} \sin{\frac{\pi(u-c)}{2w}}+\frac{1}{16}
  \sin{\frac{3 \pi(u-c)}{2w}}, & \quad \text{if } c-w < u < c+w \\
  1, & \quad \text{if } u \geq c+w.
\end{cases}
\end{equation}
We take $W_1\equiv W(u_1,1,1/4)$ with
\begin{equation}
  u_1=\frac{r_{12}+r_{34}}{2(r_{13}+r_{24})}+
  \frac{r_{12}+r_{34}}{2(r_{23}+r_{14})},
\end{equation}
where $r_{ij}$ indicates the distance between atom $i$ and $j$. Atoms 1 and 2
are the K-atoms and atoms 3 and 4 are the Na-atoms and
$W_2\equiv W(u_2,1/2,1/16)$ with
\begin{equation}
  u_2=\frac{r_{23}+r_{14}}{r_{13}+r_{24}+r_{23}+r_{14}}.
\end{equation}

\begin{figure*}
\includegraphics[height=2.4in]{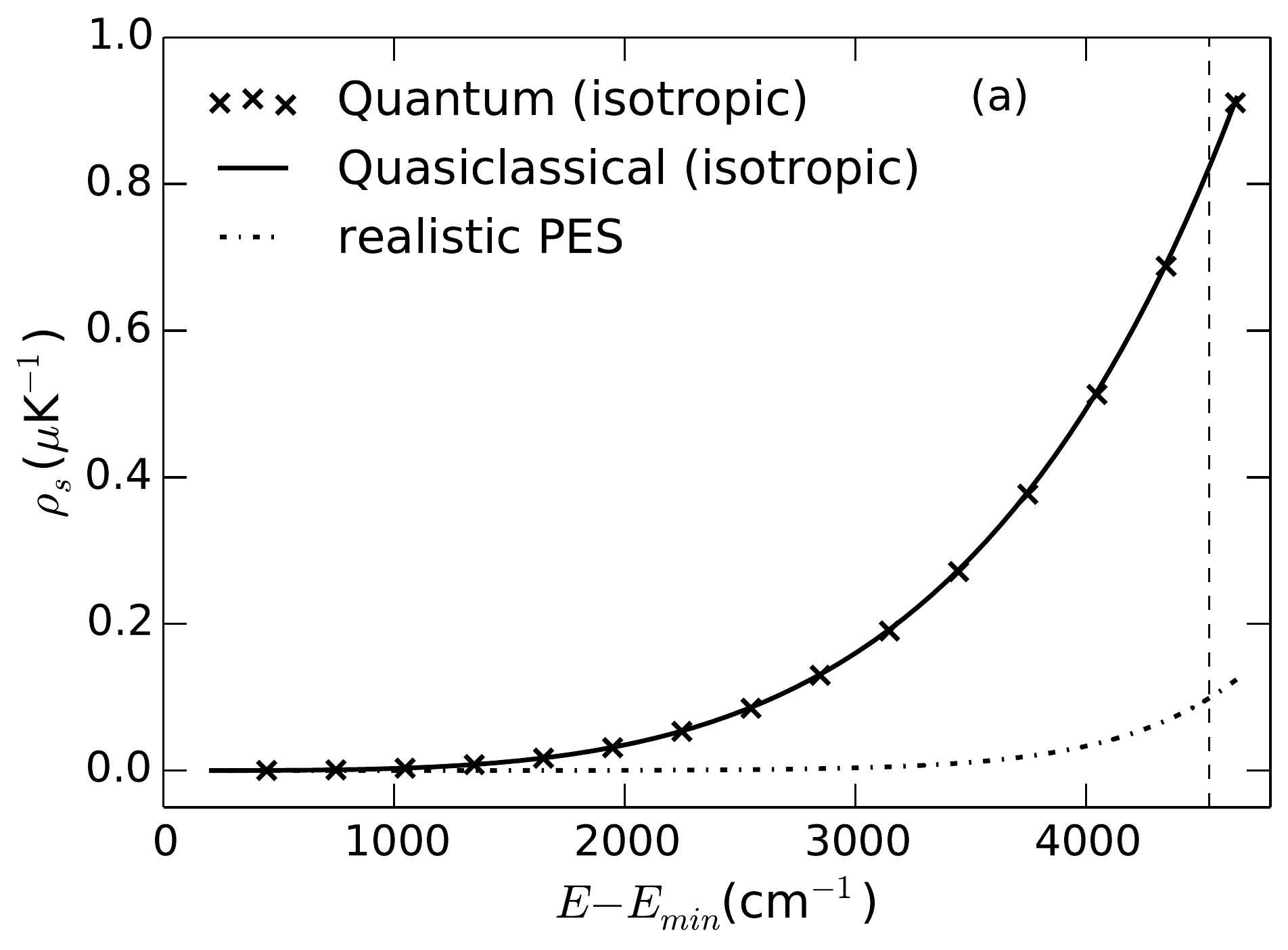} \\
\includegraphics[height=2.4in]{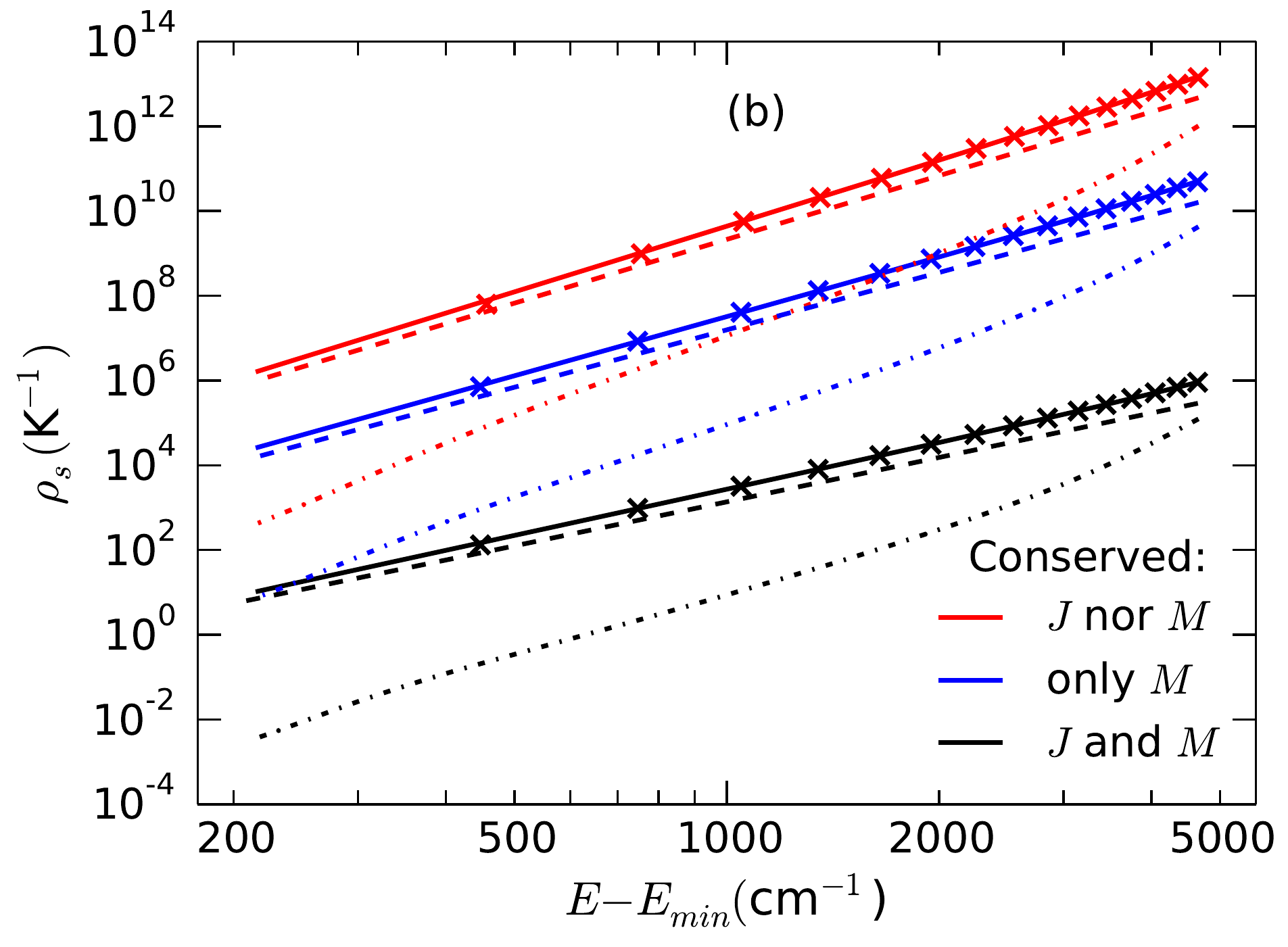} 
\captionsetup{format=plain,justification=raggedright,singlelinecheck=false}
\caption[]{The DOS of NaK+NaK as a function of the energy, $E$, for quantum
mechanical (crosses), quasiclassical (solid line) calculations for an isotropic
PES and quasiclassical calculations for a realistic PES (dashed-dotted lines).
Here $J=0$ and the quantum mechanical parity $p=1$. In panel (a) this is
plotted only for the field-free case on a linear scale, in panel (b) it is also
plotted for the cases with fields on a double logarithmic scale. 
The vertical dashed line in panel (a) indicates the classical dissociation
enery of the complex, and the dashed lines in panel (b) are straight lines
with slopes 5, 4.5, and 3.5 and only serve to illustrate the power-law energy
dependence of the DOS.}
\label{fig:fig2}
\end{figure*}

Figure \ref{fig:fig2} shows the DOS for both the isotropic $r$-independent PES
(quantum and quasiclassical) and for the realistic PES. For the isotropic PES,
the NaK monomer potentials of Ref.~\cite{christianen:2019} were used, together
with a Lennard-Jones intermolecular potential with parameters $C_6 = 8500$~$E_h
a_0^{6}$.\ and $D_e=4534$ cm$^{-1}$.
First, we note that the quasiclassical-quantum correspondence is even better
than in the case of K$_2$-Rb. This is not surprising given the DOS is larger by
about five orders of magnitude at the dissociation energy.  This results in
fewer quantum fluctuations in the DOS.
At the dissociation energy, the difference in DOS between the isotropic and
realistic PESs is about one order of magnitude, both with and without angular
momentum conservation. The slope of the DOS in Fig.~\ref{fig:fig2}(b) is much
larger and less constant for the realistic PES than for the isotropic PES. This
is due to anharmonicity and anisotropy of the PES. At the dissociation energy,
the DOS for the realistic PES is found to be (for $J=0$): 0.124 $\mu$K$^{-1}$
in the field-free case, 2.14 nK$^{-1}$ if only $M$ is conserved, and 5.12
pK$^{-1}$ when neither $J$ nor $M$ is conserved. These DOS values correspond
to RRKM sticking times of 5.96 $\mu$s, 103 ms, and 24.5 s, for these
three cases, respectively.

\subsection{Extrapolating the NaK-NaK results} \label{sec:extrapol}

In this section we estimate the DOS for other bialkali-bialkali systems by
extrapolating the accurate DOS we obtained for NaK-NaK. To find approximate
scaling laws, we use the values of $\det \mathcal{I} (\bm{y})$, $\det
\mathcal{A} (\bm{y})$ in a planar, antiparallel configuration with the bond
lengths $r_{1,2}$ at their equilibrium distances $r_0$ and the intermolecular
distance $R$ at the minimum distance of the Lennard-Jones potential
$R_0=(C_6/D_e)^{1/6}$. We assume the potential is isotropic and
harmonic, with the force constants $k_R=72 D_e/R_0^2$ and
$k_r=\omega^2 m_\mathrm{1} m_\mathrm{2}/(m_\mathrm{1}+m_\mathrm{2})$,
with $\omega$ the vibration frequency of the diatom. Furthermore, we consider
only one arrangement, meaning we drop one factor $1/2$ for the
symmetry.  Substituting this into Eq.~\eqref{eq:DOS_ABCD} yields an approximate
DOS:
\begin{equation} \label{eq:harmexpNaKNaK}
\tilde{\rho}(m_1,m_2,r_0,D_e,C_6,\omega)=\frac{256 \pi^{10} (2J+1) m_{1}^{5/2}
m_{2}^{5/2} R_0 r_0^3 }{105 h^{9} (m_{1}+m_{2})^{3/2}
\sqrt{\frac{m_1+m_{2}}{2}R_0^2+2\frac{m_{1}m_2}{m_{1}+m_2} r_0^2}}\,
\frac{1}{k_r\sqrt{k_R}}  D_e^{\frac{7}{2}} C_\mathrm{corr}.
\end{equation}
We use the DOS calculated using our realistic PES to determine the factor
$C_{\mathrm{corr}}$ in the above expression, which is meant to correct for the
anisotropy and anharmonicity of the PES. We find $C_\mathrm{corr}=0.23$.
We fix the value of this correction factor, and subsequently evaluate
Eq.~\eqref{eq:harmexpNaKNaK} for all polar bialkali-bialkali systems.  We use
diatom properties from Ref.~\cite{fedorov:2014}, and $C_6$ coefficients and
$D_e$ values from Ref.~\cite{byrd:2013}.  The resulting DOSs are listed listed
in table \ref{tab:allalkali}.  We see that---as expected from the
equations---the DOS strongly increases when moving from lighter to heavier
alkali systems.  Here, the reduced mass plays a bigger role than the total
mass, e.g., compare NaK to LiCs.  We see that the sticking times of the
collision complexes, in the absence of chemical reactions, change over three
orders of magnitude when moving from 0.25~$\mu$s for NaLi, to 253~$\mu$s for
RbCs.  Note that for fermionic molecules, $s$-wave scattering is forbidden and
that therefore $p$-wave scattering is the dominant mechanism. Therefore $J=1$ and
the sticking time is increased by a factor $2J+1=3$, see Eq.~\eqref{eq:genDOS}.

\begin{table*}[t]
\centering
\captionsetup{format=plain,justification=raggedright,singlelinecheck=false}
\caption{The estimated DOS, in $\mu$K$^{-1}$ (RRKM sticking time, in $\mu$s) for all singlet polar bialkali molecules ($J=0$) in a single hyperfine state. The star indicates the NaK-NaK sticking time has been determined accurately, without extrapolation.}
\label{tab:allalkali}
\begin{tabular}{c|cccc}
\hline
\hline
&  $^{23}$Na & $^{39}$K & $^{87}$Rb & $^{133}$Cs \\
\hline
$^{7}$Li&0.0051 (0.25)&0.014 (0.67)&0.024 (1.17)& 0.068 (3.3)\\
$^{23}$Na&.&0.124(6.0)*&0.27 (12.9)&0.83 (40)\\
$^{39}$K&.&.&0.48 (23.0)&1.50 (72)\\
$^{87}$Rb&.&.&.&5.3 (253)\\
\hline
\hline
\end{tabular}
\end{table*}

\subsection{Sticking-amplified three-body loss}
\label{sec:threebod}

We use the calculated sticking times to study one particular loss
mechanism that has been hypothesized to be responsible for the losses observed
experimentally: sticking-amplified three-body loss. Here, a free diatom
collides with a collision complex, leading to energy transfer from the complex
to the diatom and the escape of both the complex and the diatom from the trap.
To estimate the rate of this three-body loss process, we need to estimate the
rate of complex-molecule collisions and compare the resulting lifetime to the
sticking time of the complex.

The rate of complex-molecule collisions can be estimated with a quantum capture
model \cite{mayle:2013}. The only unknown parameter here is the dispersion
($C_6$) coefficient for complex-molecule collisions, which sets the mean
scattering length and rate.  This dispersion coefficient can be calculated from
the dynamic dipole polarizabilities $\alpha(i\omega)$ at imaginary frequencies
of both collision partners, A and B, using the Casimir-Polder relation,
\begin{equation} \label{eq:C6}
  C_6=\frac{3}{\pi}\int_0^\infty \alpha^\mathrm{A} (i\omega) \alpha^\mathrm{B} (i \omega)
  d\omega.
\end{equation}
Quantum mechanically, the polarizability for a given state $i$ can be
calculated from a sum over states, $f$, where $\omega_{fi}$ is the transition
frequency
\begin{equation} \label{eq:quantpolar}
  \alpha_{i}(i \omega)=\frac{2}{3} \sum_{f  \neq i} \omega_{fi} \frac{|\langle f|
  \hat{\bm{\mu}}| i \rangle |^2}{\omega_{fi}^2+\omega^2}.
\end{equation}
From Eq.\ \eqref{eq:quantpolar} it is clear that the static dipole
polarizability ($\omega=0$) is an upper limit for the polarizability. For a
ground state molecule $\omega_{fi}$ is always positive. In the case of
diatom-diatom collisions,  the dispersion coefficient is mainly due to
rotational dispersion \cite{zuchowski:13} and  $\alpha_\mathrm{diatom}(0)$ is
given approximately by
\begin{equation}
  \alpha_\mathrm{diatom}(0)=\frac{d^2}{3B},
\end{equation}
where $B$ is the rotational constant and $d$ the dipole moment. The complex is
clearly not in the ground state, meaning that terms of the sum in
Eq.~\eqref{eq:quantpolar} in energy above and below the energy level of the complex
could cancel to some extent, leading to a much smaller polarizability.
Quasiclassically, this can be quantified for the static dipole polarizability.
We derive in Appendix Sec.~\ref{appendix3} that this static dipole
polarizability can be expressed in terms of the following expectation value
\begin{equation} 
  \alpha_\mathrm{complex}(0)= \left\langle \frac{2 d^2}{3 T_\mathrm{kin}}
  \right\rangle_{\bm{q},\bm{\Omega}}.
\end{equation}
Note that here $d$ is the total dipole moment of the complex, which depends on
the geometry.  This expression is remarkably similar to the expression for the
free diatom. For the NaK-NaK system, the interaction energy can rise up to
$4534$ cm$^{-1}$ \cite{christianen:2019}.  This means that the expectation
value of $T_\mathrm{kin}$ is in the order of $10^3$ cm$^{-1}$, which is four
orders of magnitude larger than the rotational constant of NaK, which is
$0.095$ cm$^{-1}$.  This means  that the rotational dispersion contribution to
the integral in Eq.~\eqref{eq:C6}, will be much smaller than in the
diatom-diatom case. Therefore the electronic dispersion term is the most
important contribution, which can be estimated to be twice the electronic
dispersion coefficient for the diatom-diatom collisions. For NaK, this means
that the dispersion coefficient for the complex-diatom collisions, will be
$17\, 000$ $E_h a_0^6$, which is an order of magnitude smaller than the value
of $500\, 000$ $E_h a_0^6$ for diatomic collisions, which may be
counter-intuitive.

Using the multichannel quantum defect theory from Ref.~\cite{mayle:2013} and
taking the limit of $T \rightarrow 0$, this dispersion coefficient gives an
$s$-wave rate coefficient of $1.1 \cdot 10^{-10}$ cm$^{3}$s$^{-1}$. Multiplying
this by a typical density of the diatoms ($4 \cdot 10^{10}$ cm$^{-3}$
\cite{park:2015}) and taking the inverse gives the lifetime of the complex due
to three-body loss. This lifetime is given by $\tau_{3b}=0.23$~s. The sticking
time of the NaK-NaK complex for $J=1$ is approximately $18~\mu s$, so the
complex dissociates much faster than it collides with a third NaK diatom.
Therefore, sticking-amplified three-body losses are not the cause of the losses
in typical experiments\cite{park:2015,park:2017}. Accounting solely for this
loss mechanism, the lifetime of the NaK gas in the trap in the experimental
conditions would be in the order of hours\cite{park:2015,park:2017}. For the
RbCs gas of Ref.~\cite{takekoshi:2014} it would be tens of minutes. For the
NaRb gas such as reported in Ref.~\cite{ye:2018}, the loss would be on the
timescale of a minute, due to the relatively high densities.

The conclusion that three-body collisions are not the cause the experimental
losses is based on the sticking time without fields and without taking into
account hyperfine transitions. The conclusion may change in the presence of
strong electric or magnetic fields, which cause $J$ to no longer be conserved.
However, it is not clear from our calculations how strong the external fields
need to be to affect the DOS. The DOS, and therefore the sticking time, can also
strongly increase in case of hyperfine transitions of the collision complex.
However, it is not directly clear whether these occur on the timescale of the
sticking time, especially since there are no unpaired electronic spins and the
hyperfine transitions must therefore be caused by coupling to the rotational
states. The strongest hyperfine coupling is due to the nuclear quadrupole
moments interacting with the changing electric field gradients during the
collisions.  Both the inclusion of hyperfine states into the model or $J$ not
being conserved may cause the sticking times to be orders of magnitude larger
and may cause the three-body collision mechanism to be more important. 

\section{Conclusion and discussion}

We have derived a quasiclassical equation for the DOS of an ultracold, $N$-atom
collision complex, for an arbitrary PES [Eq.~\eqref{eq:genDOS}].  We have
established the accuracy of our quasiclassical method by comparing to exact
quantum results for the K$_2$-Rb and NaK-NaK system, with isotropic
$r$-independent model PESs.  We have calculated the DOS for an accurate NaK-NaK
PES to be 0.124 $\mu$K$^{-1}$, with an associated RRKM sticking time of
5.96~$\mu$s.  We extrapolate our results to the other
bialkali-bialkali systems.  The resulting DOS increases rapidly with atomic
mass, but only up to 5~$\mu$K$^{-1}$ for the heaviest system RbCs, two orders
of magnitude below what was reported previously \cite{mayle:2013}.  Using the
resulting sticking times, we conclude a sticking-amplified three-body loss mechanism
is not the cause of losses in the experiments.

\section{Acknowledgements}
We thank James Croft and John Bohn for valuable discussions and Dongzheng Yang for carefully checking all derivations in the manuscript.
T.K. is supported by NWO Rubicon grant 019.172EN.007 and an NSF grant to ITAMP.

\bibliography{manuscript}

\begin{thebibliography}{28}%
\makeatletter
\providecommand \@ifxundefined [1]{%
 \@ifx{#1\undefined}
}%
\providecommand \@ifnum [1]{%
 \ifnum #1\expandafter \@firstoftwo
 \else \expandafter \@secondoftwo
 \fi
}%
\providecommand \@ifx [1]{%
 \ifx #1\expandafter \@firstoftwo
 \else \expandafter \@secondoftwo
 \fi
}%
\providecommand \natexlab [1]{#1}%
\providecommand \enquote  [1]{``#1''}%
\providecommand \bibnamefont  [1]{#1}%
\providecommand \bibfnamefont [1]{#1}%
\providecommand \citenamefont [1]{#1}%
\providecommand \href@noop [0]{\@secondoftwo}%
\providecommand \href [0]{\begingroup \@sanitize@url \@href}%
\providecommand \@href[1]{\@@startlink{#1}\@@href}%
\providecommand \@@href[1]{\endgroup#1\@@endlink}%
\providecommand \@sanitize@url [0]{\catcode `\\12\catcode `\$12\catcode
  `\&12\catcode `\#12\catcode `\^12\catcode `\_12\catcode `\%12\relax}%
\providecommand \@@startlink[1]{}%
\providecommand \@@endlink[0]{}%
\providecommand \url  [0]{\begingroup\@sanitize@url \@url }%
\providecommand \@url [1]{\endgroup\@href {#1}{\urlprefix }}%
\providecommand \urlprefix  [0]{URL }%
\providecommand \Eprint [0]{\href }%
\providecommand \doibase [0]{http://dx.doi.org/}%
\providecommand \selectlanguage [0]{\@gobble}%
\providecommand \bibinfo  [0]{\@secondoftwo}%
\providecommand \bibfield  [0]{\@secondoftwo}%
\providecommand \translation [1]{[#1]}%
\providecommand \BibitemOpen [0]{}%
\providecommand \bibitemStop [0]{}%
\providecommand \bibitemNoStop [0]{.\EOS\space}%
\providecommand \EOS [0]{\spacefactor3000\relax}%
\providecommand \BibitemShut  [1]{\csname bibitem#1\endcsname}%
\let\auto@bib@innerbib\@empty
\bibitem [{\citenamefont {DeMille}(2002)}]{demille:2002}%
  \BibitemOpen
  \bibfield  {author} {\bibinfo {author} {\bibfnamefont {D.}~\bibnamefont
  {DeMille}},\ }\href {\doibase 10.1103/PhysRevLett.88.067901} {\bibfield
  {journal} {\bibinfo  {journal} {Phys. Rev. Lett.}\ }\textbf {\bibinfo
  {volume} {88}},\ \bibinfo {pages} {067901} (\bibinfo {year}
  {2002})}\BibitemShut {NoStop}%
\bibitem [{\citenamefont {Yelin}\ \emph {et~al.}(2006)\citenamefont {Yelin},
  \citenamefont {Kirby},\ and\ \citenamefont {C\^ot\'e}}]{yelin:2006}%
  \BibitemOpen
  \bibfield  {author} {\bibinfo {author} {\bibfnamefont {S.~F.}\ \bibnamefont
  {Yelin}}, \bibinfo {author} {\bibfnamefont {K.}~\bibnamefont {Kirby}}, \ and\
  \bibinfo {author} {\bibfnamefont {R.}~\bibnamefont {C\^ot\'e}},\ }\href
  {\doibase 10.1103/PhysRevA.74.050301} {\bibfield  {journal} {\bibinfo
  {journal} {Phys. Rev. A}\ }\textbf {\bibinfo {volume} {74}},\ \bibinfo
  {pages} {050301} (\bibinfo {year} {2006})}\BibitemShut {NoStop}%
\bibitem [{\citenamefont {Ni}\ \emph {et~al.}(2018)\citenamefont {Ni},
  \citenamefont {Rosenband},\ and\ \citenamefont {Grimes}}]{ni:2018}%
  \BibitemOpen
  \bibfield  {author} {\bibinfo {author} {\bibfnamefont {K.-K.}\ \bibnamefont
  {Ni}}, \bibinfo {author} {\bibfnamefont {T.}~\bibnamefont {Rosenband}}, \
  and\ \bibinfo {author} {\bibfnamefont {D.~D.}\ \bibnamefont {Grimes}},\
  }\href {\doibase 10.1039/C8SC02355G} {\bibfield  {journal} {\bibinfo
  {journal} {Chem. Sci.}\ }\textbf {\bibinfo {volume} {9}},\ \bibinfo {pages}
  {6830} (\bibinfo {year} {2018})}\BibitemShut {NoStop}%
\bibitem [{\citenamefont {Micheli}\ \emph {et~al.}(2006)\citenamefont
  {Micheli}, \citenamefont {Brennen},\ and\ \citenamefont
  {Zoller}}]{micheli:2006}%
  \BibitemOpen
  \bibfield  {author} {\bibinfo {author} {\bibfnamefont {A.}~\bibnamefont
  {Micheli}}, \bibinfo {author} {\bibfnamefont {G.~K.}\ \bibnamefont
  {Brennen}}, \ and\ \bibinfo {author} {\bibfnamefont {P.}~\bibnamefont
  {Zoller}},\ }\href {\doibase http://dx.doi.org/10.1038/nphys287
  10.1038/nphys287} {\bibfield  {journal} {\bibinfo  {journal} {Nat. Phys.}\
  }\textbf {\bibinfo {volume} {2}},\ \bibinfo {pages} {341} (\bibinfo {year}
  {2006})}\BibitemShut {NoStop}%
\bibitem [{\citenamefont {B\"uchler}\ \emph {et~al.}(2007)\citenamefont
  {B\"uchler}, \citenamefont {Demler}, \citenamefont {Lukin}, \citenamefont
  {Micheli}, \citenamefont {Prokof'ev}, \citenamefont {Pupillo},\ and\
  \citenamefont {Zoller}}]{buchler:2007}%
  \BibitemOpen
  \bibfield  {author} {\bibinfo {author} {\bibfnamefont {H.~P.}\ \bibnamefont
  {B\"uchler}}, \bibinfo {author} {\bibfnamefont {E.}~\bibnamefont {Demler}},
  \bibinfo {author} {\bibfnamefont {M.}~\bibnamefont {Lukin}}, \bibinfo
  {author} {\bibfnamefont {A.}~\bibnamefont {Micheli}}, \bibinfo {author}
  {\bibfnamefont {N.}~\bibnamefont {Prokof'ev}}, \bibinfo {author}
  {\bibfnamefont {G.}~\bibnamefont {Pupillo}}, \ and\ \bibinfo {author}
  {\bibfnamefont {P.}~\bibnamefont {Zoller}},\ }\href {\doibase
  10.1103/PhysRevLett.98.060404} {\bibfield  {journal} {\bibinfo  {journal}
  {Phys. Rev. Lett.}\ }\textbf {\bibinfo {volume} {98}},\ \bibinfo {pages}
  {060404} (\bibinfo {year} {2007})}\BibitemShut {NoStop}%
\bibitem [{\citenamefont {Cooper}\ and\ \citenamefont
  {Shlyapnikov}(2009)}]{cooper:2009}%
  \BibitemOpen
  \bibfield  {author} {\bibinfo {author} {\bibfnamefont {N.~R.}\ \bibnamefont
  {Cooper}}\ and\ \bibinfo {author} {\bibfnamefont {G.~V.}\ \bibnamefont
  {Shlyapnikov}},\ }\href {\doibase 10.1103/PhysRevLett.103.155302} {\bibfield
  {journal} {\bibinfo  {journal} {Phys. Rev. Lett.}\ }\textbf {\bibinfo
  {volume} {103}},\ \bibinfo {pages} {155302} (\bibinfo {year}
  {2009})}\BibitemShut {NoStop}%
\bibitem [{\citenamefont {Krems}(2008)}]{krems:08}%
  \BibitemOpen
  \bibfield  {author} {\bibinfo {author} {\bibfnamefont {R.~V.}\ \bibnamefont
  {Krems}},\ }\href {\doibase 10.1039/b802322k} {\bibfield  {journal} {\bibinfo
   {journal} {Phys. Chem. Chem. Phys.}\ }\textbf {\bibinfo {volume} {10}},\
  \bibinfo {pages} {4079} (\bibinfo {year} {2008})}\BibitemShut {NoStop}%
\bibitem [{\citenamefont {Ospelkaus}\ \emph {et~al.}(2010)\citenamefont
  {Ospelkaus}, \citenamefont {Ni}, \citenamefont {Wang}, \citenamefont
  {de~Miranda}, \citenamefont {Neyenhuis}, \citenamefont {Qu{\'e}m{\'e}ner},
  \citenamefont {Julienne}, \citenamefont {Bohn}, \citenamefont {Jin},\ and\
  \citenamefont {Ye}}]{ospelkaus:2010b}%
  \BibitemOpen
  \bibfield  {author} {\bibinfo {author} {\bibfnamefont {S.}~\bibnamefont
  {Ospelkaus}}, \bibinfo {author} {\bibfnamefont {K.-K.}\ \bibnamefont {Ni}},
  \bibinfo {author} {\bibfnamefont {D.}~\bibnamefont {Wang}}, \bibinfo {author}
  {\bibfnamefont {M.~H.~G.}\ \bibnamefont {de~Miranda}}, \bibinfo {author}
  {\bibfnamefont {B.}~\bibnamefont {Neyenhuis}}, \bibinfo {author}
  {\bibfnamefont {G.}~\bibnamefont {Qu{\'e}m{\'e}ner}}, \bibinfo {author}
  {\bibfnamefont {P.~S.}\ \bibnamefont {Julienne}}, \bibinfo {author}
  {\bibfnamefont {J.~L.}\ \bibnamefont {Bohn}}, \bibinfo {author}
  {\bibfnamefont {D.~S.}\ \bibnamefont {Jin}}, \ and\ \bibinfo {author}
  {\bibfnamefont {J.}~\bibnamefont {Ye}},\ }\href {\doibase
  10.1126/science.1184121} {\bibfield  {journal} {\bibinfo  {journal}
  {Science}\ }\textbf {\bibinfo {volume} {327}},\ \bibinfo {pages} {853}
  (\bibinfo {year} {2010})}\BibitemShut {NoStop}%
\bibitem [{\citenamefont {Andreev}\ \emph {et~al.}(2018)\citenamefont
  {Andreev}, \citenamefont {Ang}, \citenamefont {DeMille}, \citenamefont
  {Doyle}, \citenamefont {Gabrielse}, \citenamefont {Haefner}, \citenamefont
  {Hutzler}, \citenamefont {Lasner}, \citenamefont {Meisenhelder},
  \citenamefont {O’Leary}, \citenamefont {Panda}, \citenamefont {West},
  \citenamefont {West},\ and\ \citenamefont {Wu}}]{andreev:2018}%
  \BibitemOpen
  \bibfield  {author} {\bibinfo {author} {\bibfnamefont {V.}~\bibnamefont
  {Andreev}}, \bibinfo {author} {\bibfnamefont {D.~G.}\ \bibnamefont {Ang}},
  \bibinfo {author} {\bibfnamefont {D.}~\bibnamefont {DeMille}}, \bibinfo
  {author} {\bibfnamefont {J.~M.}\ \bibnamefont {Doyle}}, \bibinfo {author}
  {\bibfnamefont {G.}~\bibnamefont {Gabrielse}}, \bibinfo {author}
  {\bibfnamefont {J.}~\bibnamefont {Haefner}}, \bibinfo {author} {\bibfnamefont
  {N.~R.}\ \bibnamefont {Hutzler}}, \bibinfo {author} {\bibfnamefont
  {Z.}~\bibnamefont {Lasner}}, \bibinfo {author} {\bibfnamefont
  {C.}~\bibnamefont {Meisenhelder}}, \bibinfo {author} {\bibfnamefont {B.~R.}\
  \bibnamefont {O’Leary}}, \bibinfo {author} {\bibfnamefont {C.~D.}\
  \bibnamefont {Panda}}, \bibinfo {author} {\bibfnamefont {A.~D.}\ \bibnamefont
  {West}}, \bibinfo {author} {\bibfnamefont {E.~P.}\ \bibnamefont {West}}, \
  and\ \bibinfo {author} {\bibfnamefont {X.~A.~C.}\ \bibnamefont {Wu}},\ }\href
  {\doibase 10.1038/s41586-018-0599-8} {\bibfield  {journal} {\bibinfo
  {journal} {Nature}\ }\textbf {\bibinfo {volume} {562}},\ \bibinfo {pages}
  {355} (\bibinfo {year} {2018})}\BibitemShut {NoStop}%
\bibitem [{\citenamefont {Takekoshi}\ \emph {et~al.}(2014)\citenamefont
  {Takekoshi}, \citenamefont {Reichs\"ollner}, \citenamefont {Schindewolf},
  \citenamefont {Hutson}, \citenamefont {Sueur}, \citenamefont {Dulieu},
  \citenamefont {Ferlaino}, \citenamefont {Grimm},\ and\ \citenamefont
  {N\"agerl}}]{takekoshi:2014}%
  \BibitemOpen
  \bibfield  {author} {\bibinfo {author} {\bibfnamefont {T.}~\bibnamefont
  {Takekoshi}}, \bibinfo {author} {\bibfnamefont {L.}~\bibnamefont
  {Reichs\"ollner}}, \bibinfo {author} {\bibfnamefont {A.}~\bibnamefont
  {Schindewolf}}, \bibinfo {author} {\bibfnamefont {J.~M.}\ \bibnamefont
  {Hutson}}, \bibinfo {author} {\bibfnamefont {C.~R.~L.}\ \bibnamefont
  {Sueur}}, \bibinfo {author} {\bibfnamefont {O.}~\bibnamefont {Dulieu}},
  \bibinfo {author} {\bibfnamefont {F.}~\bibnamefont {Ferlaino}}, \bibinfo
  {author} {\bibfnamefont {R.}~\bibnamefont {Grimm}}, \ and\ \bibinfo {author}
  {\bibfnamefont {H.-C.}\ \bibnamefont {N\"agerl}},\ }\href {\doibase
  10.1103/PhysRevLett.113.205301} {\bibfield  {journal} {\bibinfo  {journal}
  {Phys. Rev. Lett.}\ }\textbf {\bibinfo {volume} {113}},\ \bibinfo {pages}
  {205301} (\bibinfo {year} {2014})}\BibitemShut {NoStop}%
\bibitem [{\citenamefont {Molony}\ \emph {et~al.}(2014)\citenamefont {Molony},
  \citenamefont {Gregory}, \citenamefont {Ji}, \citenamefont {Lu},
  \citenamefont {K\"oppinger}, \citenamefont {Sueur}, \citenamefont {Blackley},
  \citenamefont {Hutson},\ and\ \citenamefont {Cornish}}]{molony:2014}%
  \BibitemOpen
  \bibfield  {author} {\bibinfo {author} {\bibfnamefont {P.~K.}\ \bibnamefont
  {Molony}}, \bibinfo {author} {\bibfnamefont {P.~D.}\ \bibnamefont {Gregory}},
  \bibinfo {author} {\bibfnamefont {Z.}~\bibnamefont {Ji}}, \bibinfo {author}
  {\bibfnamefont {B.}~\bibnamefont {Lu}}, \bibinfo {author} {\bibfnamefont
  {M.~P.}\ \bibnamefont {K\"oppinger}}, \bibinfo {author} {\bibfnamefont
  {C.~R.~L.}\ \bibnamefont {Sueur}}, \bibinfo {author} {\bibfnamefont {C.~L.}\
  \bibnamefont {Blackley}}, \bibinfo {author} {\bibfnamefont {J.~M.}\
  \bibnamefont {Hutson}}, \ and\ \bibinfo {author} {\bibfnamefont {S.~L.}\
  \bibnamefont {Cornish}},\ }\href {\doibase 10.1103/PhysRevLett.113.255301}
  {\bibfield  {journal} {\bibinfo  {journal} {Phys. Rev. Lett.}\ }\textbf
  {\bibinfo {volume} {113}},\ \bibinfo {pages} {255301} (\bibinfo {year}
  {2014})}\BibitemShut {NoStop}%
\bibitem [{\citenamefont {Guo}\ \emph {et~al.}(2016)\citenamefont {Guo},
  \citenamefont {Zhu}, \citenamefont {Lu}, \citenamefont {Ye}, \citenamefont
  {Wang}, \citenamefont {Vexiau}, \citenamefont {Bouloufa-Maafa}, \citenamefont
  {Qu\'em\'ener}, \citenamefont {Dulieu},\ and\ \citenamefont
  {Wang}}]{guo:2016}%
  \BibitemOpen
  \bibfield  {author} {\bibinfo {author} {\bibfnamefont {M.}~\bibnamefont
  {Guo}}, \bibinfo {author} {\bibfnamefont {B.}~\bibnamefont {Zhu}}, \bibinfo
  {author} {\bibfnamefont {B.}~\bibnamefont {Lu}}, \bibinfo {author}
  {\bibfnamefont {X.}~\bibnamefont {Ye}}, \bibinfo {author} {\bibfnamefont
  {F.}~\bibnamefont {Wang}}, \bibinfo {author} {\bibfnamefont {R.}~\bibnamefont
  {Vexiau}}, \bibinfo {author} {\bibfnamefont {N.}~\bibnamefont
  {Bouloufa-Maafa}}, \bibinfo {author} {\bibfnamefont {G.}~\bibnamefont
  {Qu\'em\'ener}}, \bibinfo {author} {\bibfnamefont {O.}~\bibnamefont
  {Dulieu}}, \ and\ \bibinfo {author} {\bibfnamefont {D.}~\bibnamefont
  {Wang}},\ }\href {\doibase 10.1103/PhysRevLett.116.205303} {\bibfield
  {journal} {\bibinfo  {journal} {Phys. Rev. Lett.}\ }\textbf {\bibinfo
  {volume} {116}},\ \bibinfo {pages} {205303} (\bibinfo {year}
  {2016})}\BibitemShut {NoStop}%
\bibitem [{\citenamefont {Park}\ \emph {et~al.}(2015)\citenamefont {Park},
  \citenamefont {Will},\ and\ \citenamefont {Zwierlein}}]{park:2015}%
  \BibitemOpen
  \bibfield  {author} {\bibinfo {author} {\bibfnamefont {J.~W.}\ \bibnamefont
  {Park}}, \bibinfo {author} {\bibfnamefont {S.~A.}\ \bibnamefont {Will}}, \
  and\ \bibinfo {author} {\bibfnamefont {M.~W.}\ \bibnamefont {Zwierlein}},\
  }\href {\doibase 10.1103/PhysRevLett.114.205302} {\bibfield  {journal}
  {\bibinfo  {journal} {Phys. Rev. Lett.}\ }\textbf {\bibinfo {volume} {114}},\
  \bibinfo {pages} {205302} (\bibinfo {year} {2015})}\BibitemShut {NoStop}%
\bibitem [{\citenamefont {See\ss{}elberg}\ \emph {et~al.}(2018)\citenamefont
  {See\ss{}elberg}, \citenamefont {Buchheim}, \citenamefont {Lu}, \citenamefont
  {Schneider}, \citenamefont {Luo}, \citenamefont {Tiemann}, \citenamefont
  {Bloch},\ and\ \citenamefont {Gohle}}]{seesselberg:2018}%
  \BibitemOpen
  \bibfield  {author} {\bibinfo {author} {\bibfnamefont {F.}~\bibnamefont
  {See\ss{}elberg}}, \bibinfo {author} {\bibfnamefont {N.}~\bibnamefont
  {Buchheim}}, \bibinfo {author} {\bibfnamefont {Z.-K.}\ \bibnamefont {Lu}},
  \bibinfo {author} {\bibfnamefont {T.}~\bibnamefont {Schneider}}, \bibinfo
  {author} {\bibfnamefont {X.-Y.}\ \bibnamefont {Luo}}, \bibinfo {author}
  {\bibfnamefont {E.}~\bibnamefont {Tiemann}}, \bibinfo {author} {\bibfnamefont
  {I.}~\bibnamefont {Bloch}}, \ and\ \bibinfo {author} {\bibfnamefont
  {C.}~\bibnamefont {Gohle}},\ }\href {\doibase 10.1103/PhysRevA.97.013405}
  {\bibfield  {journal} {\bibinfo  {journal} {Phys. Rev. A}\ }\textbf {\bibinfo
  {volume} {97}},\ \bibinfo {pages} {013405} (\bibinfo {year}
  {2018})}\BibitemShut {NoStop}%
\bibitem [{\citenamefont {Park}\ \emph {et~al.}(2017)\citenamefont {Park},
  \citenamefont {Yan}, \citenamefont {Loh}, \citenamefont {Will},\ and\
  \citenamefont {Zwierlein}}]{park:2017}%
  \BibitemOpen
  \bibfield  {author} {\bibinfo {author} {\bibfnamefont {J.~W.}\ \bibnamefont
  {Park}}, \bibinfo {author} {\bibfnamefont {Z.~Z.}\ \bibnamefont {Yan}},
  \bibinfo {author} {\bibfnamefont {H.}~\bibnamefont {Loh}}, \bibinfo {author}
  {\bibfnamefont {S.~A.}\ \bibnamefont {Will}}, \ and\ \bibinfo {author}
  {\bibfnamefont {M.~W.}\ \bibnamefont {Zwierlein}},\ }\href {\doibase
  10.1126/science.aal5066} {\bibfield  {journal} {\bibinfo  {journal}
  {Science}\ }\textbf {\bibinfo {volume} {357}},\ \bibinfo {pages} {372}
  (\bibinfo {year} {2017})}\BibitemShut {NoStop}%
\bibitem [{\citenamefont {Ye}\ \emph {et~al.}(2018)\citenamefont {Ye},
  \citenamefont {Guo}, \citenamefont {Gonz{\'a}lez-Mart{\'\i}nez},
  \citenamefont {Qu{\'e}m{\'e}ner},\ and\ \citenamefont {Wang}}]{ye:2018}%
  \BibitemOpen
  \bibfield  {author} {\bibinfo {author} {\bibfnamefont {X.}~\bibnamefont
  {Ye}}, \bibinfo {author} {\bibfnamefont {M.}~\bibnamefont {Guo}}, \bibinfo
  {author} {\bibfnamefont {M.~L.}\ \bibnamefont {Gonz{\'a}lez-Mart{\'\i}nez}},
  \bibinfo {author} {\bibfnamefont {G.}~\bibnamefont {Qu{\'e}m{\'e}ner}}, \
  and\ \bibinfo {author} {\bibfnamefont {D.}~\bibnamefont {Wang}},\ }\href@noop
  {} {\bibfield  {journal} {\bibinfo  {journal} {Sci. Adv.}\ }\textbf {\bibinfo
  {volume} {4}} (\bibinfo {year} {2018})}\BibitemShut {NoStop}%
\bibitem [{\citenamefont {Qu\'em\'ener}\ and\ \citenamefont
  {Julienne}(2012)}]{quemener:2012}%
  \BibitemOpen
  \bibfield  {author} {\bibinfo {author} {\bibfnamefont {G.}~\bibnamefont
  {Qu\'em\'ener}}\ and\ \bibinfo {author} {\bibfnamefont {P.~S.}\ \bibnamefont
  {Julienne}},\ }\href {\doibase 10.1021/cr300092g} {\bibfield  {journal}
  {\bibinfo  {journal} {Chem. Rev.}\ }\textbf {\bibinfo {volume} {112}},\
  \bibinfo {pages} {4949} (\bibinfo {year} {2012})}\BibitemShut {NoStop}%
\bibitem [{\citenamefont {Mayle}\ \emph {et~al.}(2013)\citenamefont {Mayle},
  \citenamefont {Qu\'em\'ener}, \citenamefont {Ruzic},\ and\ \citenamefont
  {Bohn}}]{mayle:2013}%
  \BibitemOpen
  \bibfield  {author} {\bibinfo {author} {\bibfnamefont {M.}~\bibnamefont
  {Mayle}}, \bibinfo {author} {\bibfnamefont {G.}~\bibnamefont {Qu\'em\'ener}},
  \bibinfo {author} {\bibfnamefont {B.~P.}\ \bibnamefont {Ruzic}}, \ and\
  \bibinfo {author} {\bibfnamefont {J.~L.}\ \bibnamefont {Bohn}},\ }\href
  {\doibase 10.1103/PhysRevA.87.012709} {\bibfield  {journal} {\bibinfo
  {journal} {Phys. Rev. A}\ }\textbf {\bibinfo {volume} {87}},\ \bibinfo
  {pages} {012709} (\bibinfo {year} {2013})}\BibitemShut {NoStop}%
\bibitem [{\citenamefont {Gregory}\ \emph {et~al.}(2019)\citenamefont
  {Gregory}, \citenamefont {Frye}, \citenamefont {Blackmore}, \citenamefont
  {Bridge}, \citenamefont {Sawant}, \citenamefont {Hutson},\ and\ \citenamefont
  {Cornish}}]{gregory:2019}%
  \BibitemOpen
  \bibfield  {author} {\bibinfo {author} {\bibfnamefont {P.~D.}\ \bibnamefont
  {Gregory}}, \bibinfo {author} {\bibfnamefont {M.~D.}\ \bibnamefont {Frye}},
  \bibinfo {author} {\bibfnamefont {J.~A.}\ \bibnamefont {Blackmore}}, \bibinfo
  {author} {\bibfnamefont {E.~M.}\ \bibnamefont {Bridge}}, \bibinfo {author}
  {\bibfnamefont {R.}~\bibnamefont {Sawant}}, \bibinfo {author} {\bibfnamefont
  {J.~M.}\ \bibnamefont {Hutson}}, \ and\ \bibinfo {author} {\bibfnamefont
  {S.~L.}\ \bibnamefont {Cornish}},\ }\href@noop {} {\bibfield  {journal}
  {\bibinfo  {journal} {preprint arXiv:1904.00654}\ } (\bibinfo {year}
  {2019})}\BibitemShut {NoStop}%
\bibitem [{\citenamefont {F.~E.~Croft}\ \emph {et~al.}(2017)\citenamefont
  {F.~E.~Croft}, \citenamefont {Balakrishnan},\ and\ \citenamefont
  {K.~Kendrick}}]{croft:2017}%
  \BibitemOpen
  \bibfield  {author} {\bibinfo {author} {\bibfnamefont {J.}~\bibnamefont
  {F.~E.~Croft}}, \bibinfo {author} {\bibfnamefont {N.}~\bibnamefont
  {Balakrishnan}}, \ and\ \bibinfo {author} {\bibfnamefont {B.}~\bibnamefont
  {K.~Kendrick}},\ }\href {\doibase 10.1103/PhysRevA.96.062707} {\bibfield
  {journal} {\bibinfo  {journal} {Phys. Rev. A}\ }\textbf {\bibinfo {volume}
  {96}},\ \bibinfo {pages} {062707} (\bibinfo {year} {2017})}\BibitemShut
  {NoStop}%
\bibitem [{\citenamefont {Mayle}\ \emph {et~al.}(2012)\citenamefont {Mayle},
  \citenamefont {Ruzic},\ and\ \citenamefont {Bohn}}]{mayle:2012}%
  \BibitemOpen
  \bibfield  {author} {\bibinfo {author} {\bibfnamefont {M.}~\bibnamefont
  {Mayle}}, \bibinfo {author} {\bibfnamefont {B.~P.}\ \bibnamefont {Ruzic}}, \
  and\ \bibinfo {author} {\bibfnamefont {J.~L.}\ \bibnamefont {Bohn}},\ }\href
  {\doibase 10.1103/PhysRevA.85.062712} {\bibfield  {journal} {\bibinfo
  {journal} {Phys. Rev. A}\ }\textbf {\bibinfo {volume} {85}},\ \bibinfo
  {pages} {062712} (\bibinfo {year} {2012})}\BibitemShut {NoStop}%
\bibitem [{\citenamefont {Levine}(2005)}]{levine:2005}%
  \BibitemOpen
  \bibfield  {author} {\bibinfo {author} {\bibfnamefont {R.~D.}\ \bibnamefont
  {Levine}},\ }\href {\doibase 10.1017/CBO9780511614125} {\emph {\bibinfo
  {title} {Molecular Reaction Dynamics}}}\ (\bibinfo  {publisher} {Cambridge
  University Press},\ \bibinfo {year} {2005})\BibitemShut {NoStop}%
\bibitem [{\citenamefont {Croft}\ \emph {et~al.}(2017)\citenamefont {Croft},
  \citenamefont {Makrides}, \citenamefont {Li}, \citenamefont {Petrov},
  \citenamefont {Kendrick}, \citenamefont {Balakrishnan},\ and\ \citenamefont
  {Kotochigova}}]{croft:2017a}%
  \BibitemOpen
  \bibfield  {author} {\bibinfo {author} {\bibfnamefont {J.~F.~E.}\
  \bibnamefont {Croft}}, \bibinfo {author} {\bibfnamefont {C.}~\bibnamefont
  {Makrides}}, \bibinfo {author} {\bibfnamefont {M.}~\bibnamefont {Li}},
  \bibinfo {author} {\bibfnamefont {A.}~\bibnamefont {Petrov}}, \bibinfo
  {author} {\bibfnamefont {B.~K.}\ \bibnamefont {Kendrick}}, \bibinfo {author}
  {\bibfnamefont {N.}~\bibnamefont {Balakrishnan}}, \ and\ \bibinfo {author}
  {\bibfnamefont {S.}~\bibnamefont {Kotochigova}},\ }\href@noop {} {\bibfield
  {journal} {\bibinfo  {journal} {Nat. Comm.}\ }\textbf {\bibinfo {volume}
  {8}},\ \bibinfo {pages} {15897} (\bibinfo {year} {2017})}\BibitemShut
  {NoStop}%
\bibitem [{\citenamefont {Christianen}\ \emph {et~al.}(2019)\citenamefont
  {Christianen}, \citenamefont {Karman}, \citenamefont {Vargas-Hern\'andez},
  \citenamefont {Groenenboom},\ and\ \citenamefont {Krems}}]{christianen:2019}%
  \BibitemOpen
  \bibfield  {author} {\bibinfo {author} {\bibfnamefont {A.}~\bibnamefont
  {Christianen}}, \bibinfo {author} {\bibfnamefont {T.}~\bibnamefont {Karman}},
  \bibinfo {author} {\bibfnamefont {R.~A.}\ \bibnamefont {Vargas-Hern\'andez}},
  \bibinfo {author} {\bibfnamefont {G.~C.}\ \bibnamefont {Groenenboom}}, \ and\
  \bibinfo {author} {\bibfnamefont {R.~V.}\ \bibnamefont {Krems}},\ }\href
  {\doibase 10.1063/1.5082740} {\bibfield  {journal} {\bibinfo  {journal} {J.
  Chem. Phys.}\ }\textbf {\bibinfo {volume} {150}},\ \bibinfo {pages} {064106}
  (\bibinfo {year} {2019})}\BibitemShut {NoStop}%
\bibitem [{\citenamefont {Peslherbe}\ and\ \citenamefont
  {Hase}(1994)}]{peslherbe:1994}%
  \BibitemOpen
  \bibfield  {author} {\bibinfo {author} {\bibfnamefont {G.~H.}\ \bibnamefont
  {Peslherbe}}\ and\ \bibinfo {author} {\bibfnamefont {W.~L.}\ \bibnamefont
  {Hase}},\ }\href {\doibase 10.1063/1.468114} {\bibfield  {journal} {\bibinfo
  {journal} {J. Chem. Phys.}\ }\textbf {\bibinfo {volume} {101}},\ \bibinfo
  {pages} {8535} (\bibinfo {year} {1994})}\BibitemShut {NoStop}%
\bibitem [{\citenamefont {Fedorov}\ \emph {et~al.}(2014)\citenamefont
  {Fedorov}, \citenamefont {Derevianko},\ and\ \citenamefont
  {Varganov}}]{fedorov:2014}%
  \BibitemOpen
  \bibfield  {author} {\bibinfo {author} {\bibfnamefont {D.~A.}\ \bibnamefont
  {Fedorov}}, \bibinfo {author} {\bibfnamefont {A.}~\bibnamefont {Derevianko}},
  \ and\ \bibinfo {author} {\bibfnamefont {S.~A.}\ \bibnamefont {Varganov}},\
  }\href {\doibase 10.1063/1.4875038} {\bibfield  {journal} {\bibinfo
  {journal} {J. Chem. Phys.}\ }\textbf {\bibinfo {volume} {140}},\ \bibinfo
  {pages} {184315} (\bibinfo {year} {2014})}\BibitemShut {NoStop}%
\bibitem [{\citenamefont {Byrd}(2013)}]{byrd:2013}%
  \BibitemOpen
  \bibfield  {author} {\bibinfo {author} {\bibfnamefont {J.}~\bibnamefont
  {Byrd}},\ }\emph {\bibinfo {title} {Ultracold Chemistry of Alkali
  Clusters}},\ \href@noop {} {Ph.D. thesis},\ \bibinfo  {school} {University of
  Connecticut} (\bibinfo {year} {2013})\BibitemShut {NoStop}%
\bibitem [{\citenamefont {\.{Z}uchowski}\ \emph {et~al.}(2013)\citenamefont
  {\.{Z}uchowski}, \citenamefont {Kosicki}, \citenamefont {Kodrycka},\ and\
  \citenamefont {Sold\'an}}]{zuchowski:13}%
  \BibitemOpen
  \bibfield  {author} {\bibinfo {author} {\bibfnamefont {P.~S.}\ \bibnamefont
  {\.{Z}uchowski}}, \bibinfo {author} {\bibfnamefont {M.}~\bibnamefont
  {Kosicki}}, \bibinfo {author} {\bibfnamefont {M.}~\bibnamefont {Kodrycka}}, \
  and\ \bibinfo {author} {\bibfnamefont {P.}~\bibnamefont {Sold\'an}},\ }\href
  {\doibase 10.1103/PhysRevA.87.022706} {\bibfield  {journal} {\bibinfo
  {journal} {Phys. Rev. A}\ }\textbf {\bibinfo {volume} {87}},\ \bibinfo
  {pages} {022706} (\bibinfo {year} {2013})}\BibitemShut {NoStop}%
\end{thebibliography}%

\newpage

\section{Appendix} 

\subsection{DOS in presence of a field} \label{appendix}

We are interested in the DOS in the presence of, e.g., an electric field,
where $J$ is no longer conserved, but $M$ still is.
For the DOS calculation this means that we can no longer treat space
isotropically and neglect the $J$-dependence of the kinetic energy. We
modify Eq.~\eqref{eq:genPSV} accordingly, resulting in
\begin{equation}
N= \frac{C_{\bm{Nm}}}{\Gamma(\frac{D}{2}+1)} \int\!  d\bm{q} \int\! d\bm{\Omega}
\sin(\beta) \int\! d\bm{J}\, G(\bm{q})
    \left\{\pi \left[E-V(\bm{q}) -\frac{\bm{J}^T \mathcal{R}(\bm{\Omega}) [\mathcal{I}(\bm{q})^{-1}]^{(\mathrm{bf})} \mathcal{R}(\bm{\Omega})^{-1} \bm{J}}{2}\right]\right\}^{\frac{D}{2}} .
\end{equation}
For conserved $M=0$, we integrate over $J_z$ from $-1/2$ to
$1/2$. This is inaccurate only for the $J=0$ state, but the
contribution of $J=0$ to the total DOS is very small if all $J$ are accessible.
Because $M_z=0$ we can neglect the kinetic energy associated with $J_z$, so
\begin{equation}
   N=\frac{2C_{\bm{Nm}}}{\Gamma(\frac{D}{2}+2)} \int\!  d\bm{q} \int\! d\bm{\Omega}
   \sin(\beta) \frac{G(\bm{q})}{\sqrt{M_{zz}}} \\
   \left\{\pi \left[E-V(\bm{q}) \right]\right\}^{\frac{D}{2}+1},
\end{equation}
where $M_{zz}$ is the minor of the $z,z$ element
of $\mathcal{R}(\bm{\Omega})^{-1} \mathcal{I}(\bm{q})^{-1} \mathcal{R}(\bm{\Omega})$.
By Cramer's rule the minor $M_{zz}$ of a matrix $\mcX$ is equal
to $(\mcX^{-1})_{zz} \det \mcX$, so
\begin{equation}
   \frac{1}{\sqrt{M_{zz}}} =
      \frac{\det \mathcal{I}(\bm{q})}
      {\sqrt{[\mathcal{R}(\bm{\Omega})^{-1} \mathcal{I}(\bm{q}) \mathcal{R}(\bm{\Omega}]_{zz}}}.
\end{equation}
The denominator in this expression depends only on the first two Euler
angles $\alpha$ and $\beta$ and we find
\begin{multline}
  \label{eq:A3}
  N=\frac{4 \pi C_{\bm{Nm}}}{\Gamma(\frac{D}{2}+2)} \int\!  d\bm{q}\, G(\bm{q})
  \sqrt{\det{\mathcal{I}(\bm{q})}} \left\{\pi
  \left[E-V(\bm{q})\right]\right\}^{\frac{D}{2}+1}  \\ \int_0^{2\pi}\! d\alpha
  \int_0^\pi\! d\beta \frac{\sin(\beta)   }{\sqrt{\mathcal{I}_1(\bm{q})
  \cos^2(\beta) + \mathcal{I}_2(\bm{q}) \sin^2(\beta) \cos^2(\alpha) +
  \mathcal{I}_3(\bm{q}) \sin^2(\beta) \sin^2(\alpha)}},
\end{multline}
where we have chosen the lower integration bound, the zeroes of $\bm{\Omega}$,
such that the inertial tensor is diagonal and the eigenvalues are ordered in
magnitude. The variables $\mathcal{I}_1(\bm{q})$, $\mathcal{I}_2(\bm{q})$, and
$\mathcal{I}_3(\bm{q})$ are the eigenvalues of $\mathcal{I}(\bm{q})$, where
$\mathcal{I}_1(\bm{q})$ is the largest and $\mathcal{I}_3(\bm{q})$ the
smallest.  This choice is possible since we integrate over all angles, such
that the integral is independent of the starting point.

The integral over $\beta$ results in
\begin{equation}
  \frac{4 \pi}{\sqrt{\mathcal{I}}_\mathrm{rot}}= \frac{1}{\sqrt{\mathcal{I}_1}}
  \int_0^{2\pi}\! d\alpha\, \frac{ \{ \log[1+f(\alpha)]-\log[1-f(\alpha)] \}}{f(\alpha)},
\end{equation}
where 
\begin{equation}
  f(\alpha)=\sqrt{1-\frac{\mathcal{I}_2 \cos^2(\alpha)+ \mathcal{I}_3
  \sin^2(\alpha)}{\mathcal{I}_1}}, 
\end{equation}
and $\mathcal{I}_\mathrm{rot}$ is the ``rotationally averaged'' value of
$\mathcal{I}_{zz}$. An analytical expression for this integral can be obtained
by expanding the logarithms as a power series. Only even powers of $f(\alpha)$
remain and all resulting integrals can be calculated analytically, yielding
\begin{align}
  \frac{4\pi}{\sqrt{\mathcal{I}}_\mathrm{rot}}&= \frac{2}{\sqrt{\mathcal{I}_1}}
  \sum_{n=0}^{\infty} \frac{1}{2n+1} \int_0^{2 \pi}
  \left[(1-\frac{\mathcal{I}_3}{\mathcal{I}_1}
  )-(\frac{\mathcal{I}_2-\mathcal{I}_3}{\mathcal{I}_1}) \cos^2(\alpha)\right]^n
  d\alpha \\
  	&= \frac{4 \pi}{\sqrt{\mathcal{I}_1}} \sum_{n=0}^{\infty}
  \frac{\left(1-\frac{\mathcal{I}_3}{\mathcal{I}_1} \right)^{n}}{2n+1}
  {}_2F_1\left(\frac{1}{2},-n;1;\frac{\mathcal{I}_2-\mathcal{I}_3}{\mathcal{I}_1-\mathcal{I}_3}\right),
\end{align}
where ${}_2F_1{}$ is the hypergeometric function.
The sum converges rapidly as long as ${\mathcal{I}_3}$ is of the same order as
$\mathcal{I}_1$ and $\mathcal{I}_2$. The values this sum can assume lie between
$1$ and $\pi/2$.  If we substitute this result into Eq.~\eqref{eq:A3},
we obtain
\begin{equation}
  N=\frac{16 \pi^2 C_{\bm{Nm}}}{\Gamma(\frac{D}{2}+2)} \int\!  d\bm{q}\, G(\bm{q})
  \sqrt{\frac{\det \mathcal{I} (\bm{q})}{\mathcal{I}_\mathrm{rot}(\bm{q})}} \{\pi
  [E-V(\bm{q}) ]\}^{\frac{D}{2}+1},
\end{equation}
and
\begin{equation}
  \rho=\frac{16 \pi^3 C_{\bm{Nm}}}{\Gamma(\frac{D}{2}+1)} \int\!  d\bm{q}\,
  G(\bm{q}) \sqrt{\frac{\det \mathcal{I}
  (\bm{q})}{\mathcal{I}_\mathrm{rot}(\bm{q})}} \{\pi [E-V(\bm{q})
  ]\}^{\frac{D}{2}}. 
\end{equation}

\subsection{Example calculations for K$_2$-Rb and NaK-NaK} \label{appendix2}

For K$_2$-Rb the kinetic energy (for $J=0$) can be written as:
\begin{equation}
  E_{kin}=\frac{\mu_{K_2Rb}}{2}\dot{R}^2+\frac{m_\mathrm{K}}{4}\dot{r}^2+\frac{m_\mathrm{K}}{4}r^2
  \dot{\theta}^2- \frac{\bm{j}^T\mathcal{I}^{-1} \bm{j}}{2}.
\end{equation}
If we define the $\bm{x}^{\mathrm{(bf)}}$ coordinates by choosing the K$_2$
molecule to be in the $xy$-plane and the $R$ to be along the $x$-axis, then
$\bm{j}$ is given by
\begin{equation}
\bm{j}=\begin{pmatrix}
0 \\
0 \\
	\frac{m_\mathrm{K}}{2} r^2 \dot{\theta}
\end{pmatrix},
\end{equation}
and $\mathcal{I}$ is given by
\begin{align*}
	\mathcal{I}_{xx}&=\frac{m_\mathrm{K}}{2} r^2 \sin^2(\theta), \\
	\mathcal{I}_{yy}&=\frac{m_\mathrm{K}}{2} r^2 \cos^2(\theta)+\mu_{K_2 Rb} R^2,   \\
	\mathcal{I}_{zz}&=\frac{m_\mathrm{K}}{2} r^2 +\mu_{K_2 Rb} R^2,   \\
	\mathcal{I}_{xy}&= \mathcal{I}_{yx}=-\frac{m_\mathrm{K}}{2} r^2 \cos(\theta) \sin(\theta), \\
\mathcal{I}_{xz} &= \mathcal{I}_{zx}= 0, \\
\mathcal{I}_{yz} &= \mathcal{I}_{zy}= 0.
\end{align*}

For the NaK-NaK system, the kinetic energy (for $J=0$) can be written as
\begin{equation}
  T_{\mathrm{kin}}=\frac{\mu_\mathrm{NaK}}{2}\left[\dot{r}_1^2+\dot{r}_2^2+r_1^2
  \dot{\theta}_1^2+r_2^2\dot{\theta}_2^2+r_2^2 \sin^2(\theta_2)
  \dot{\phi}^2\right]+\frac{m_\mathrm{Na}+m_\mathrm{K}}{4}
  \dot{R}^2-\frac{\bm{j}^T \mathcal{I}^{-1} \bm{j}}{2}.
\end{equation}
If we choose $r_1$ and $R$ to lie in the $xy$-plane, with $R$ along the $x$-axis,
then $\bm{j}$ is given by
\begin{equation}
  \bm{j}=\begin{pmatrix}
  \mu_\mathrm{NaK} r_2^2 \sin^2(\theta_2) \dot{\phi} \\
  -\mu_\mathrm{NaK} [r_2^2 \sin(\phi) \dot{\theta_2}+r_2^2 \sin(\theta_2) \cos(\theta_2) \cos(\phi) \dot{\phi}] \\
  \mu_\mathrm{NaK} [r_1^2 \dot{\theta_1}+r_2^2 \cos(\phi) \dot{\theta_2}-r_2^2 \sin(\theta_2) \cos(\theta_2) \sin(\phi) \dot{\phi}]
  \end{pmatrix}.
\end{equation}
For the elements of $\mathcal{I}$ we find
\begin{align*}
\mathcal{I}_{xx}&=\mu_\mathrm{NaK}[r_1^2 \sin^2(\theta_1)+r_2^2 \sin^2(\theta_2)], \\
	\mathcal{I}_{yy}&=\mu_\mathrm{NaK}[r_1^2 \cos^2(\theta_1)+r_2^2 \cos^2(\theta_2)+r_2^2 \sin^2(\theta_2) \sin^2(\phi)]+\frac{m_\mathrm{K}+m_\mathrm{Na}}{2}R^2,   \\
	\mathcal{I}_{zz}&= \mu_\mathrm{NaK} [r_1^2+r_2^2 \cos^2(\theta_2)+r_2^2 \sin^2(\theta_2) \cos^2(\phi)] +\frac{m_\mathrm{K}+m_\mathrm{Na}}{2}R^2,   \\
\mathcal{I}_{xy}&=\mathcal{I}_{yx}=-\mu_\mathrm{NaK}[r_1^2 \cos(\theta_1) \sin(\theta_1)+r_2^2 \cos(\theta_2) \sin(\theta_2) \cos(\phi)], \\
\mathcal{I}_{xz} &= \mathcal{I}_{zx}= -\mu_\mathrm{NaK} r_2^2 \cos(\theta_2) \sin(\theta_2) \sin(\phi), \\
\mathcal{I}_{yz} &= \mathcal{I}_{zy}= -\mu_\mathrm{NaK} r_2^2 \sin^2(\theta_2)  \sin(\phi)\cos(\phi).
\end{align*}

\subsection{Polarizability of a complex} \label{appendix3}
In Sec.~\ref{sec:threebod} we argue qualitatively that the
static dipole polarizability of the complex is much smaller than for the
diatoms. Here we use our quasiclassical formalism to express the polarizability
as an expectation value over phase space. For the complex, we can calculate
the expectation value of the static dipole polarizability in our quasiclassical
framework.  If we re-introduce the integral over the Euler angles $\bm{\Omega}$
in Eq.~\eqref{eq:genDOS} and introduce an external field $\bm{F}$, we can write
the DOS as
\begin{equation}
  \rho(\bm{F})=\int d\bm{q}\, d\bm{\Omega}\, \sin(\beta)\, \delta
  \rho(\bm{q},\bm{\Omega},\bm{F}), 
\end{equation}
where $\delta \rho(\bm{q},\bm{\Omega})$ is given by
\begin{equation}
  \delta \rho(\bm{q},\bm{\Omega},\bm{F})=\frac{g_{\bm{N}Jp}\pi^{1+\frac{D}{2}}
  \hbar^3 C_{\bm{Nm}} (2J+1)}{\Gamma(\frac{D}{2}) }  G(\bm{q})
  [E-V(\bm{q})-f(\bm{q},\bm{\Omega},\bm{F})]^{\frac{D}{2}-1}.
\end{equation}

Here $f(\bm{q},\bm{\Omega},\bm{F})$ is a perturbation on the energy caused by
the external field $\bm{F}$, which we assume to be small enough for
$\mathcal{J}$ to still be (approximately) conserved.  Then the expectation of
$\chi(\bm{q},\bm{\Omega})$ can be calculated as
\begin{equation} \label{eq:expval}
  \langle \chi \rangle_{\bm{q},\bm{\Omega}}(\bm{F}) = \frac{1}{\rho} \int\!
  d\bm{q}\, d\bm{\Omega}\, \sin(\beta)\, \delta \rho(\bm{q},\bm{\Omega},\bm{F})\,
  \chi(\bm{q},\bm{\Omega},\bm{F}).
\end{equation}

The polarizability tensor $\alpha$ is given by
\begin{equation}
  \alpha_{ij}=\frac{\partial d_i}{\partial \mathcal{E}_j},
\end{equation}
where $\bm{\mathcal{E}}$ is the electric field and $\bm{d}(\bm{q},\bm{\Omega})$
the electric dipole moment.  The electric dipole moment is given by the vector
sum of the dipoles of the two NaK molecules.  These molecular dipoles lie along
the molecular axes, the directions of which depend on $\bm{q}$ and
$\bm{\Omega}$.

The interaction energy of the system with an electric field is given by
$-\bm{d} \cdot \bm{\mathcal{E}}$.  The expectation value of the electric dipole
moment, for a given $J$, is given by
\begin{equation}
  \langle d_i \rangle_{\bm{q},\bm{\Omega}} (\bm{\mathcal{E}}) =
  \frac{g_{\bm{N}Jp}\pi^{1+\frac{D}{2}} \hbar^3 C_{\bm{Nm}}
  (2J+1)}{\Gamma(\frac{D}{2}) \rho} \int\! d\bm{q}\, d\bm{\Omega}\, \sin(\beta)\,
  G(\bm{q}) [E-V(\bm{q})+\bm{d}(\bm{q},\bm{\Omega}) \cdot
  \bm{\mathcal{E}}]^{\frac{D}{2}-1}d_i(\bm{q},\bm{\Omega})  .
\end{equation} 

Because of the integration over $\bm{\Omega}$, the expectation value of the
dipole moment in the weak field limit vanishes.  For the polarizability, the
off-diagonal components integrate to zero, but the diagonal components do not.
The expectation values of the diagonal polarizability components (if we take
$\bm{\mathcal{E}}= 0$) are given by
\begin{equation}
   \langle \alpha_{ii} \rangle_{\bm{q},\bm{\Omega}} = \frac{g_{\bm{N}Jp}
  \pi^{1+\frac{D}{2}} \hbar^3 C_{\bm{Nm}} (2J+1)}{\Gamma(\frac{D}{2}-1) \rho}
  \int\! d\bm{q}\, d\bm{\Omega}\, \sin(\beta)\, G(\bm{q})
  [E-V(\bm{q})]^{\frac{D}{2}-2}d_i(\bm{q},\bm{\Omega})^2  .
\end{equation} 
If we introduce the isotropic polarizability as
$\alpha_0=\frac{1}{3}(\alpha_{xx}+\alpha_{yy}+\alpha_{zz})$, we obtain
\begin{equation}
   \langle \alpha_0 \rangle_{\bm{q},\bm{\Omega}} = \frac{g_{\bm{N}Jp}
  \pi^{1+\frac{D}{2}} \hbar^3 C_{\bm{Nm}} (2J+1)}{3 \Gamma(\frac{D}{2}-1)}
   \int\!
  d\bm{q}\, d\bm{\Omega}\, \sin(\beta)\, G(\bm{q})
  [E-V(\bm{q})]^{\frac{D}{2}-2}d(\bm{q})^2  .
\end{equation}
Comparing to Eq.~\eqref{eq:expval}, this expression can be written as
\begin{equation}
  \langle \alpha_0 \rangle_{\bm{q},\bm{\Omega}} = \frac{\frac{D}{2}-1}{3 \rho}
  \int\! d\bm{q}\, d\bm{\Omega}\, \sin(\beta)
  \frac{\delta\rho(\bm{q},\bm{\Omega},0)d(\bm{q})^2}{E-V(\bm{q})}
  = \frac{\frac{D}{2}-1}{3} \left\langle \frac{d^2}{T_\mathrm{kin}}
  \right\rangle_{\bm{q},\bm{\Omega}}.
\end{equation}
For a diatom-diatom complex $D=6$ and therefore the static dipole
polarizability becomes
\begin{equation} 
  \alpha_0= \left\langle \frac{2 d^2}{3 T_\mathrm{kin}}
\right\rangle_{\bm{q},\bm{\Omega}}.
\end{equation}

\end{document}